\documentclass[journal]{IEEEtran}
\usepackage{amsmath,amsfonts}

\usepackage{algorithm}
\usepackage{algorithmic}
\usepackage{array}
\usepackage[caption=false,font=footnotesize,labelfont=rm,textfont=rm]{subfig}
\usepackage{textcomp}
\usepackage{stfloats}
\usepackage{url}
\usepackage{verbatim}
\usepackage{graphicx}
\usepackage{cite}
\usepackage{threeparttable}
\usepackage{makecell}
\usepackage{booktabs}
\usepackage{multirow}
\usepackage{bbding}

\begin{document}

\title{Aqua-Sim Fourth Generation: Towards General and Intelligent Simulation for Underwater Acoustic Networks}

\author{Jiani~Guo,
    Shanshan~Song*,~\IEEEmembership{Member,~IEEE,}
    Hao~Chen,
    Bingwen~Huangfu,
	Jun~Liu,	
	and~Jun-Hong~Cui
	\IEEEcompsocitemizethanks{
		\IEEEcompsocthanksitem Shanshan Song is the corresponding author. 
		\IEEEcompsocthanksitem Jiani Guo, Shanshan Song, Hao Chen, and Bingwen Huangfu are with the College of Computer Science and Technology, Jilin University, Changchun 130012, China (e-mail: jnguo20@mails.jlu.edu.cn; songss@jlu.edu.cn; haochen22@mails.jlu.edu.cn; hfbw24@mails.jlu.edu.cn).
		\IEEEcompsocthanksitem Jun Liu is with the School of Electronic and Information Engineering, Beihang University, Beijing 100191, China (e-mail: liujun2019@buaa.edu.cn).		
		\IEEEcompsocthanksitem Jun-Hong Cui is with the College of Computer Science and Technology, Jilin University, Changchun 130012, China, and also with the Shenzhen Ocean Information Technology Industry Research Institute, Shenzhen 518055, China (e-mail: junhong\_cui@jlu.edu.cn).}
}

\markboth{Journal of \LaTeX\ Class Files,~Vol.~14, No.~8, August~2021}%
{Shell \MakeLowercase{\textit{et al.}}: A Sample Article Using IEEEtran.cls for IEEE Journals}

\maketitle
\begin{abstract}
Simulators are essential to troubleshoot and optimize Underwater Acoustic Network (UAN) schemes (network protocols and communication technologies) before real field experiments. However, due to programming differences between the above two contents, most existing simulators concentrate on one while weakening the other, leading to non-generic simulations and biased performance results. Moreover, novel UAN schemes increasingly integrate Artificial Intelligence (AI) techniques, yet existing simulators lack support for necessary AI frameworks, failing to train and evaluate these intelligent methods. On the other hand, these novel schemes consider more UAN characteristics involving more complex parameter configurations, which also challenges simulators in flexibility and fineness. To keep abreast of advances in UANs, we propose the Fourth Generation (FG) ns-3-based simulator Aqua-Sim~FG, enhancing the general and intelligent simulation ability. On the basis of retaining previous generations' functions, we design a new general architecture, which is compatible with various programming languages, including MATLAB, C++, and Python. In this way, Aqua-Sim~FG provides a general environment to simulate communication technologies, network protocols, and AI models simultaneously. In addition, we expand six new features from node and communication levels by considering the latest UAN methods' requirements, which enhances the simulation flexibility and fineness of Aqua-Sim~FG. Experimental results show that Aqua-Sim~FG can simulate UANs' performance realistically, reflect intelligent methods' problems in real-ocean scenarios, and provide more effective troubleshooting and optimization for actual UANs. The basic simulator is available at https://github.com/JLU-smartocean/aqua-sim-fg.
\end{abstract}

\begin{IEEEkeywords}
Underwater acoustic network, Underwater network simulation, Network troubleshooting and optimization.
\end{IEEEkeywords}

\section{Introduction}
\IEEEPARstart{S}{imulators} can verify the effectiveness of whole networking schemes (communication technologies and network protocols) in low-cost and controlled environments \cite{r1}. This is especially critical for Underwater Acoustic Networks (UANs), which face complex deployment and expensive upkeep costs in real field experiments. However, due to unique UAN characteristics, such as non-standardized protocols, changeable channel conditions, and long propagation delay, terrestrial network simulators fail to simulate UAN schemes directly. Therefore, researchers constantly propose and improve special simulators to effectively evaluate UAN schemes before real field experiments.

Early studies focus on constructing ocean environment databases and designing sound filed models, aiming to provide a near-real environment for a single-layer scheme simulation \cite{r2, r3, r4}. With in-depth studies, researchers gradually develop complete systems to simulate full-layer UAN schemes. In this process, Discrete Event Simulation (DES) \cite{r5, r6, r7} is recognized as a predominant paradigm for UANs, which executes an event (e.g., send or receive packet) at a particular instant in the simulation timeline and changes the network state. Under such a condition, users can collect packet traces on each node or link to deeply understand the UAN scheme. Leveraging these advantages, researchers design DES-based simulators to simulate accurate packet-by-packet traces across various UAN scenarios with different schemes. To further simplify operations and increase scalability, these simulators are continually improved with a broad range of specialized features \cite{r8, r9, r10}. Although existing UAN simulators make some advances in authenticity and functionality, they still face the following issues as UANs develop.

\textbf{Separate simulation of network and communication.} UANs' communication technologies and channel models are generally simulated by MATLAB since they involve complex data analysis and matrix calculation. However, existing DES-based simulators (ns-2 or ns-3) implement UAN protocols' simulations via C++, which struggles with handling complex mathematical calculations \cite{r5} \cite{r11}. Therefore, the communication process is usually simplified or even ignored in existing UAN simulators. Such separated simulations fail to reflect interactions between network and communication, resulting in biased troubleshooting and optimization.

\textbf{Incompatible with intelligent methods.} Artificial Intelligence (AI) techniques are promising for UANs' signal process, multiple underwater vehicles' path planning, collaborative resource allocation, etc., since they adapt to dynamic underwater environments and excel at discovering trends behind complex data \cite{r12, r13, r14}. However, most core AI packages are implemented by Python. DES-based simulators are incompatible with these packages, failing to train AI algorithms. Similarly, Python lacks an event-driven framework like DES, limiting its ability to simulate UAN schemes. The current solution is to simulate AI-based algorithms solely via Python without considering UAN characteristics \cite{r15} \cite{r16}. As a result, existing simulations fail to reflect the actual effects of underwater environments on AI methods' performance.       
 
\textbf{Inflexible and coarse-grained configuration.} Increasing demands of underwater applications introduce more and more complex UAN studies, such as cross-layer protocols \cite{r17}, \cite{r18}, multi-dimensional resource allocation \cite{r13}, and heterogeneous node cooperation \cite{r19}. These novel studies require flexible and refined configurations of node and communication resources. However, existing simulators generally provide fixed and coarse-grained configurations, increasing the difficulty of these studies' simulations.  

To keep abreast of advances in UANs, we propose the Fourth Generation (FG) ns-3-based simulator Aqua-Sim~FG, providing general and intelligent simulation for UANs. In Aqua-Sim~FG, we design a new general and intelligent architecture on the basis of retaining previous generations' functions. The architecture integrates different programming languages, including MATLAB, C++, and Python. In this way, Aqua-Sim~FG can simulate communication technologies, network protocols, and AI models simultaneously under a unified environment, reflecting their real interactions. Moreover, we expand six new features of node and communication in Aqua-Sim~FG to simulate the latest UAN schemes conveniently. For node features, we 1) consider the movement characteristics and design different mobility models for typical underwater vehicles; 2) reconstruct header structure to adjust the header length based on specific packet types and reduce storage space adaptively; 3) introduce a new tailer structure to assist in exchanging information between layers and decrease cross-layer interaction complexity. For communication features, we 1) modify the multi-channel model to support the configuration of subcarrier-level frequency spectrum; 2) introduce multiple modulation models to provide multi-rate simulation; 3) improve propagation models to simulate channel attenuation with different accuracy degrees. Based on the above new features, Aqua-Sim~FG enhances simulations' flexibility and fineness.  

We summarize our contributions as follows:

\begin{itemize}
    \item We propose a new general and intelligent architecture in Aqua-Sim~FG, which integrates MATLAB, C++, and Python to simulate communication technologies, network protocols, and AI models simultaneously, assisting users in analyzing their mutual influences under a near-real ocean scenario.
    \item We expand six new features from node and communication levels, involving various mobility models, adaptive header structure, simple cross-layer interaction, subcarrier-level frequency spectrum configuration, multiple modulation models, and different accuracy propagation models, to enhance the UAN simulations' flexibility and fineness. 
    \item We compare Aqua-Sim~FG with real field experiments, and the comparison results show that Aqua-Sim~FG can simulate UANs' performance realistically. Compared with the previous generation simulator, Aqua-Sim~FG can reflect intelligent methods' problems in real-ocean scenarios and provide more effective troubleshooting and optimization for actual UANs.  
\end{itemize}

\section{Related Work}

\subsection{Communication and Network Simulators}

MATLAB is a mainstream simulator for communication technologies. It provides extensive function libraries and toolboxes to achieve signal processing, encoding and decoding, modulation, and demodulation. Moreover, MATLAB can interface with the hardware to realize hardware-in-the-loop simulation, improving the simulation's authenticity. 

As critical tools for network practitioners, DES-based simulators have existed for decades \cite{r20}. Notable examples include ns-2, ns-3, and OPNET \cite{r5, r6, r11}. OPNET adopts a three-level modeling mechanism: process level, node level, and network level. The process level simulates the behavior of a single object, then the node level interconnects objects into a device, and finally, the network level combines these devices to form a network. However, OPNET is a commercial software that can not be used for free. Ns-2 and ns-3 are two open-source simulators that support users to expand and change functions based on their own requirements. Due to the diversity of UAN studies, such self-developed simulators are more suitable for UAN simulations.

\subsection{Underwater Acoustic Network Simulators}

According to the above analysis, most existing UAN simulators are designed based on ns-2 and ns-3. \cite{r21} and \cite{r22} proposed two ns-2-based UAN simulators with supporting libraries from ns-2-Miracle. \cite{r2} designed a framework, called World Ocean Simulation System (WOSS), which focuses on simulating acoustic propagation of underwater networks, supporting libraries from ns-2-Miracle as well as Bellhop (a sound propagation simulator based on ray tracing). \cite{r10} proposed the first generation simulator (Aqua-Sim) of Aqua-Sim series, following the object-oriented design of ns-2 and implementing a complete protocol stack from the physical layer to the application layer.

Although Aqua-Sim provides relatively complete UAN simulation, it also suffers from many limitations such as poorly arranged architecture, steep user learning curve, restricted real-system module support, and poor memory performance. Moreover, ns-2 has its limitations and was gradually superseded by ns-3. Therefore, \cite{r23} proposed the Next Generation (NG) simulator based on ns-3, called Aqua-Sim~NG. Compared with Aqua-Sim, Aqua-Sim~NG based on ns-3 uses smart pointers, which automatically deallocate objects through reference counting to reduce memory leaks. In addition, Aqua-Sim~NG reconstructs the architecture of Aqua-Sim and provides better modularization management of simulator's protocol layers. To improve the UAN simulation authenticity, \cite{r24} proposed the third generation (TG) simulator, called Aqua-Sim~TG, which integrates Bellhop and real underwater acoustic modems, achieving semi-physical simulation based on real marine information. To date, the first three generations of the Aqua-Sim series have been downloaded more than 7,000 times.  

However, previous generations of simulators based on C++ fail to simulate communication technologies based on MATLAB and AI methods based on Python. Moreover, they generally adopt fixed and coarse-grained configurations that can not simulate complex UAN studies with fine-grained configuration requirements. In this paper, we propose the fourth generation simulator Aqua-Sim~FG, which integrates MATLAB, C++, and Python to achieve general and intelligent UAN simulations. Moreover, we expand six new features from node and communication levels to enhance flexibility and fineness of the simulator configuration. In this way, Aqua-Sim~FG improves the ability of troubleshooting and optimization for real UANs.   

\begin{figure*}[!t]
	\centering
	\includegraphics[width=17cm]{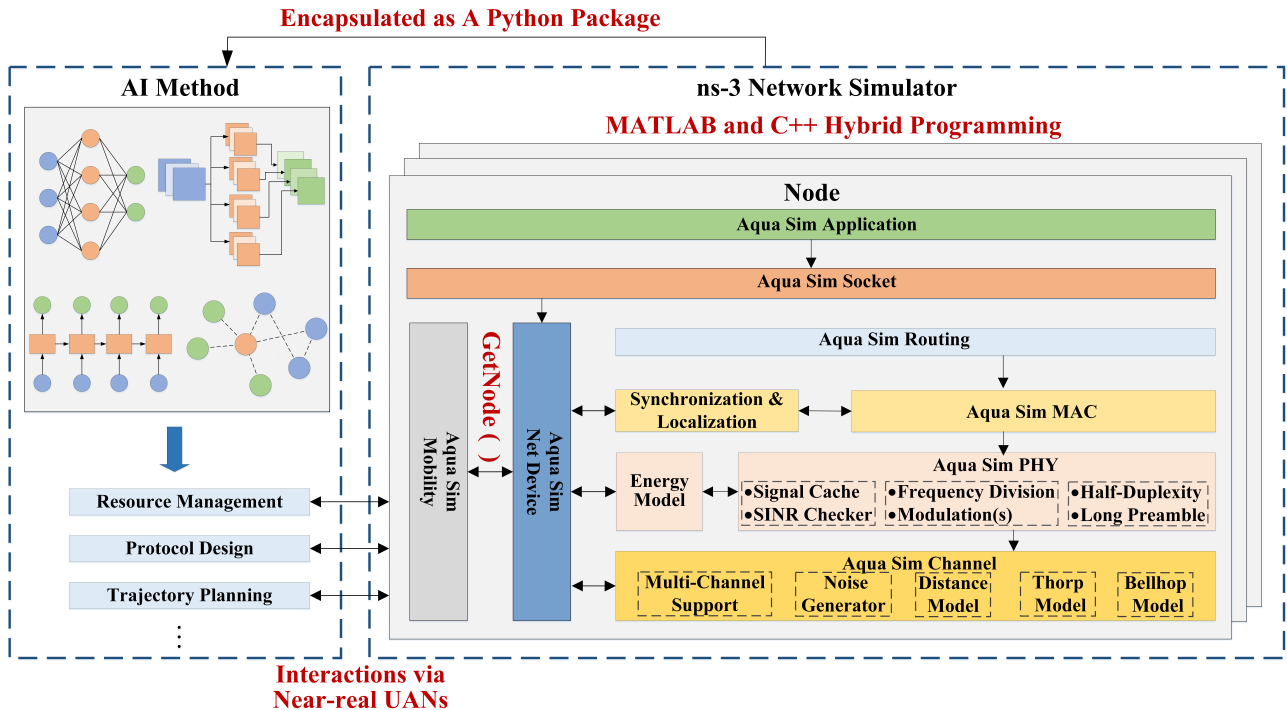}
	\caption{Aqua-Sim fourth generation architecture.}
	\label{p3.1}
\end{figure*}

\section{Architecture Overview}
Fig.~\ref{p3.1} depicts the overall architecture of the proposed Aqua-Sim~FG simulator. As an expansion of Aqua-Sim~TG, Aqua-Sim~FG can still be used as an independent UANs' simulator based on ns-3, simulating multiple underwater communication nodes. Users can configure different scale UANs with various topologies and design abundant network protocols to conduct expansive simulation tests. Each node in a virtual UAN maintains its own device information and protocol stack. In the channel layer, the node employs the device list information to communicate with other nodes. It is worth noting that although device information of these nodes can be set individually (such as mobility models, positions, IDs, and packets), network-related information should be configured uniformly (such as protocols of the routing layer, Medium Access Control (MAC) layer, and physical layer, as well as the channel layer's propagation model), to ensure nodes can reliably establish a network as real marine scenarios.         
  
To integrate communication technology and network protocols in a unified simulation environment, we achieve MATLAB and C++ hybrid programming in Aqua-Sim~FG, decreasing the development difficulty of numerical analysis and matrix operations in a single C++ environment. In this way, Aqua-Sim~FG reflects interaction effects between communication technology and network protocols, reducing the development costs for users. 

Moreover, employing AI tools has become a significant trend in UAN research. However, most existing AI-related packages and frameworks are designed based on Python. To detailed study the performance of AI methods in UANs, Aqua-Sim~FG encapsulates the whole virtual UAN based on C++ as a Python Package. Therefore, an AI algorithm employs the package to rebuild a virtual UAN in the Python environment. Compared with developing solely in Python, Aqua-Sim~FG provides communication-related information that is much closer to real UANs for underwater AI research. 

Besides the above major improvements, we expand six new features from node and communication levels, achieving mobile UAN simulation, adaptive packet header adjustment, flexible layer interaction, subcarrier-level frequency spectrum configuration, multi-rate setting, and different accuracy signal propagation. Based on these new features, flexibility and fineness are improved in this generation simulator.   

 \begin{figure}[t]
	\centering	

	\subfloat[Configure MATLAB-related information in \tt{CMakeLists.txt} \label{p3.2a}]{\includegraphics[width=8cm]{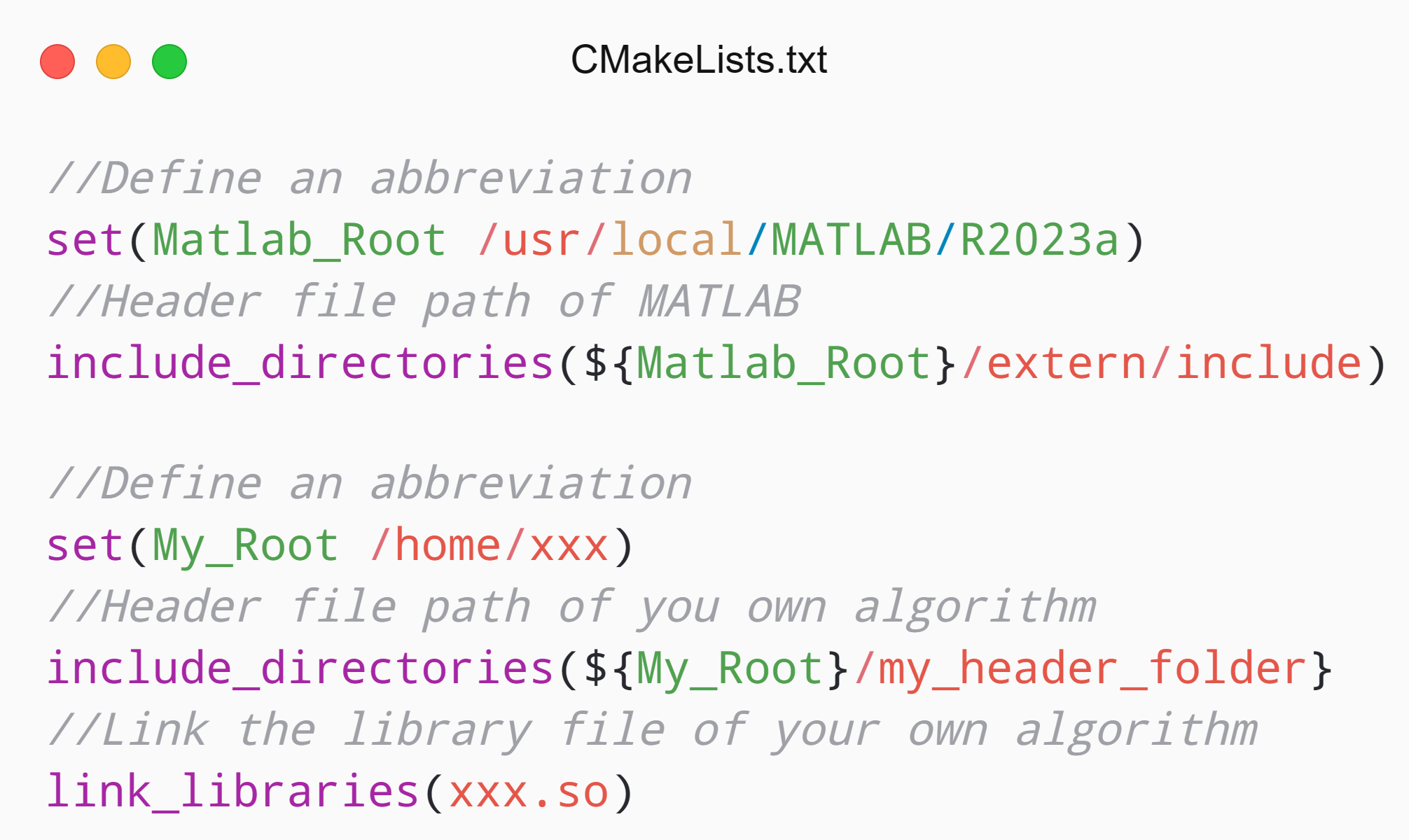}}

	\subfloat[An example of calling a MATLAB function in a C++ file \label{p3.2b}]{\includegraphics[width=8cm ]{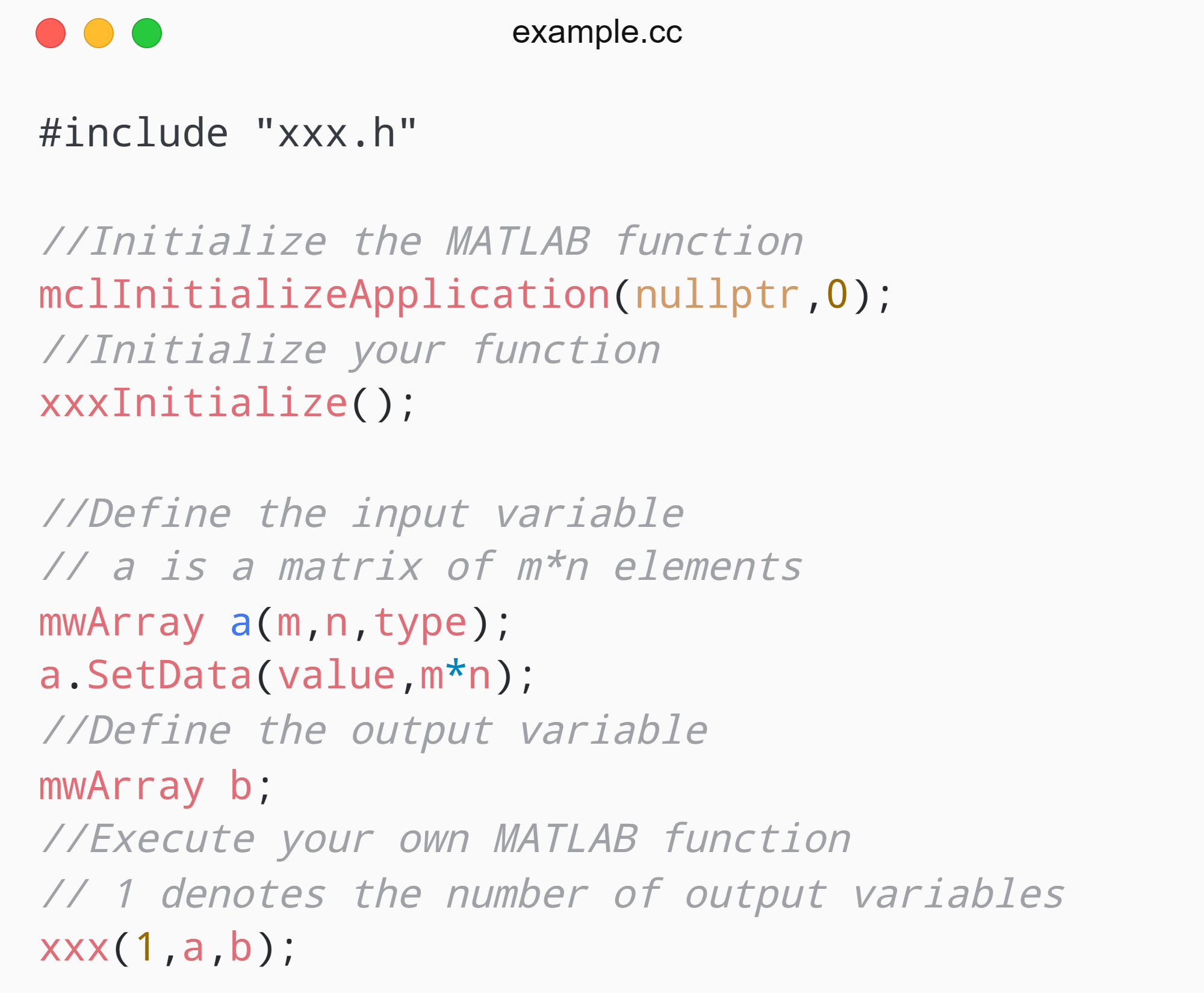}}					
	\caption{The concrete implementation of MATLAB and C++ hybrid programming.}
	\label{p3.2}	

 \vspace{-0.5cm}
\end{figure} 

\section{New Features}
\subsection{MATLAB and C++ Hybrid Programming}   

To decrease the complexity of code developing and running, Aqua-Sim~FG supports C++ to call MATLAB functions. In the first use of Aqua-Sim~FG, users need to install MATLAB Runtime matching the MATLAB version. The MATLAB Runtime is a standalone set of shared libraries that enables the execution of compiled MATLAB, Simulink applications, or components. After the installation, we append library paths of MATLAB Runtime to {\tt{LD$\_$LIBRARY$\_$PATH}} environment variable based on the tooltip.

Second, we compile a custom MATLAB function {\tt{xxx()}} to a C++ library and configure related information in the ns-3 environment. Compared with Aqua-Sim~TG, which can only be used in ns-3.27, we update Aqua-Sim~FG in a higher ns-3 version (ns-3.38) to prevent our simulator from not adapting to the current version libraries. In ns-3.38, the configuration file is called {\tt{CMakeLists.txt}}, and we append the MATLAB-related header files' paths and the compiled library {\tt{xxx.so}} to this file, as represented in Fig. \ref{p3.2a}. In this way, we can call the specific MATLAB function in a C++ environment. Take Fig.~\ref{p3.2b} as an example, a C++ source file should first execute MATLAB-related initialization functions. Then, we define input and output variables based on {\tt{xxx()}} function format. Finally, we can obtain the value of output variables in the ns-3 environment by calling the MATLAB function. After calculating the required values via MATLAB functions, Aqua-Sim~FG further utilizes these values in C++ programming to simulate UANs.    

\subsection{UAN Simulations with AI Methods}  

To explore the real performance of AI methods in UANs, Aqua-Sim~FG integrates virtual UANs and AI tools in the same scenario. For this function, the core idea is to encapsulate the C++ UAN as a Python package, enabling the AI framework to obtain UAN information from the DES-based simulator. During the specific UAN simulations based on AI methods, users should first install {\tt{cppyy}}, a flexible tool for automatic Python-C++ bindings without any language extensions or intermediate languages. After the C++ encapsulation, we can call all UAN information in a Python scenario. 

\begin{figure}[!t]
	\centering
	\includegraphics[width=8cm]{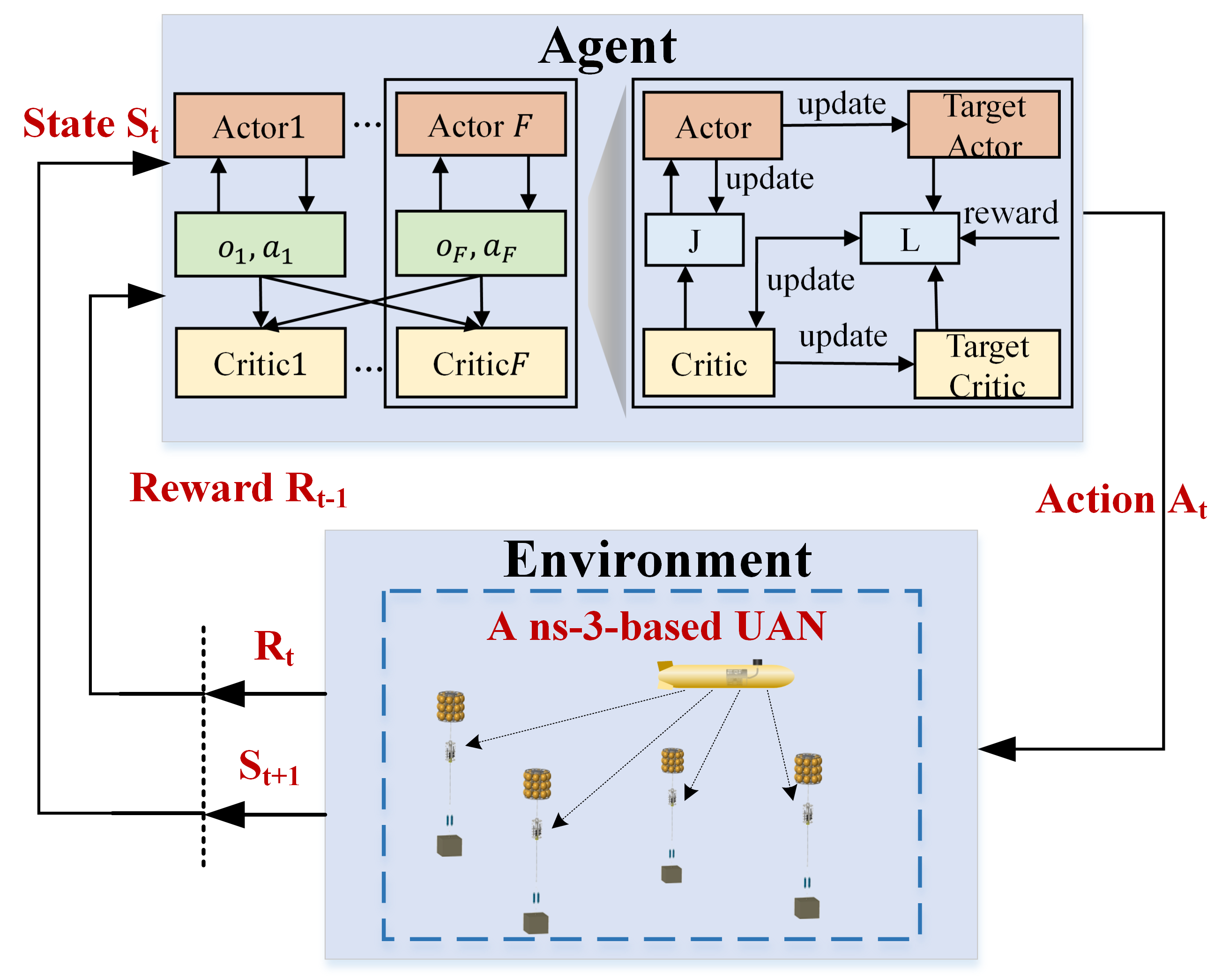}
	\caption{An example of a reinforcement learning framework in Aqua-Sim~FG. The ns-3-based UAN is encapsulated as a Python package called {\tt{ns}}. By employing this package, the agent calls a specific node in an ns-3-based UAN to execute an action.}
	\label{p3.3}
\end{figure}

As represented in Fig.~\ref{p3.3}, we take the Reinforcement Learning (RL) as an example of AI methods. Traditional UAN simulations with RL methods generally artificially assume the communication results to obtain states, failing to reflect unique UAN characteristics in the training process. In Aqua-Sim~FG, an agent in RL employs the encapsulated C++ package {\tt{ns}} to call {\tt{ns3::NodeContainer}} and access the specific node in a UAN. Then, the node executes an action determined by the RL method, and the ns-3-based UAN simulates the effects of this action on the reward and UAN state. Aqua-Sim~FG enables AI methods to train in a scenario that replicates the unique UAN characteristics, helping to enhance the practicality of underwater AI methods.      

\subsection{New Node Features} 

\subsubsection{Mobility Models of Typical Underwater Vehicles}
\begin{figure*}[t]
	\centering	

	\subfloat[The zigzag movement of an UG]{\includegraphics[width=4.6cm]{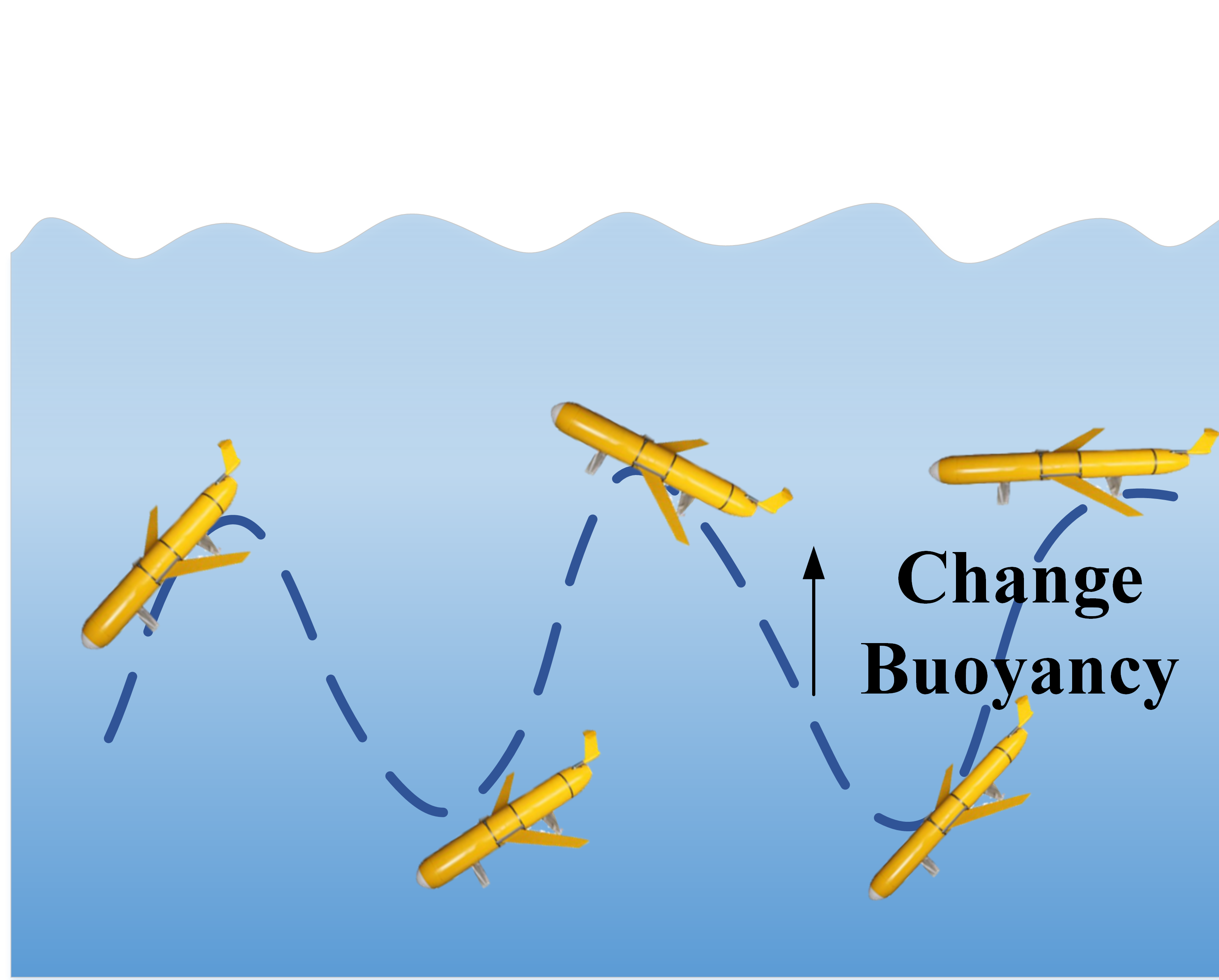}\label{p3.5a}}	
	\hspace{0.5cm}
	\subfloat[The wave movement of a WG]{\includegraphics[width=5cm ]{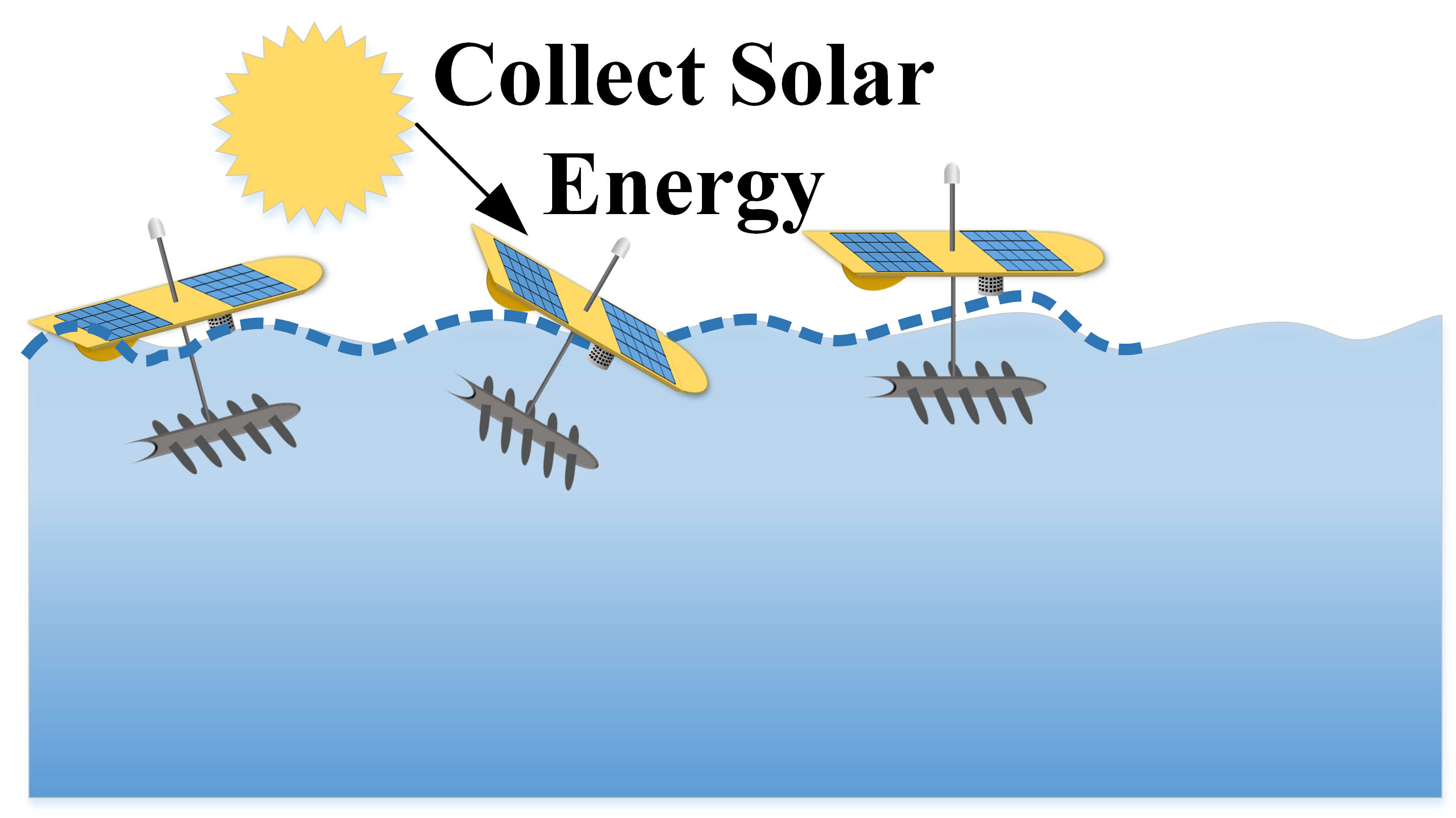}\label{p3.5c}}	
 	\hspace{0.5cm}
 	\subfloat[The preset trajectory of an AUV]{\includegraphics[width=4.7cm ]{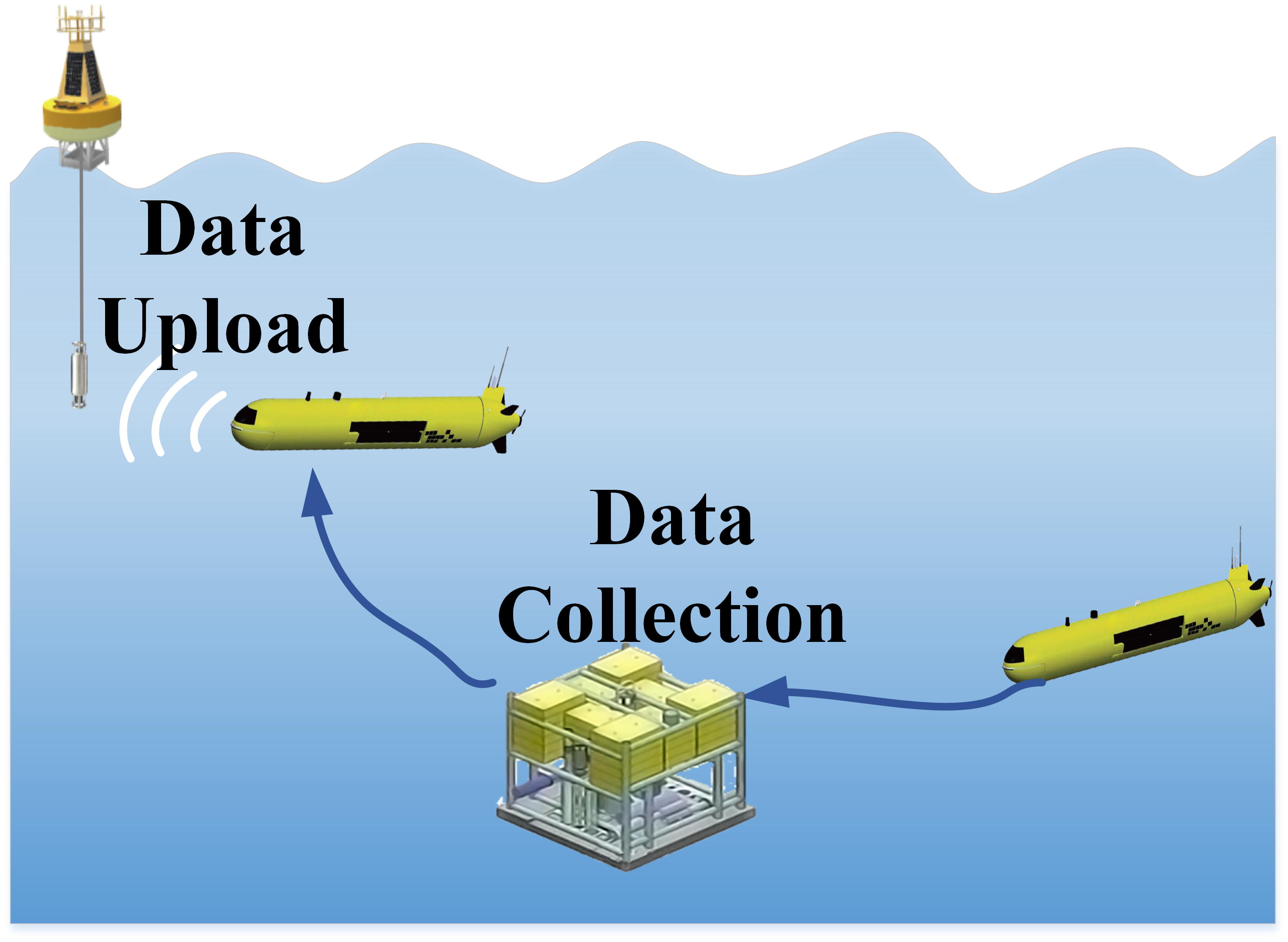}\label{p3.5b}}		
	\caption{Movement characteristics of typical underwater vehicles.}
	\label{p3.5}	
\end{figure*}

As underwater vehicles gradually develop, devices in real UANs are more and more abundant. To better simulate the movement of different underwater vehicles in UANs, Aqua-Sim~FG develops three new mobility models by considering the movement characteristics of typical vehicles, including Underwater Glider (UG), Wave Glider (WG), and Autonomous Underwater Vehicle (AUV), as represented in Fig.~\ref{p3.5}.

UG adjusts buoyancy to change its depth and employs the hydrodynamics of fixed wings to achieve underwater zigzag gliding motion. In the system, the movement trajectory is simulated by the {\tt{Aqua-Sim~FG::UG Mobility}} module. Each underwater glider is implemented as a node and binds with the above mobility module. Before the simulation starts, the user needs to preset the velocity, glide depth range, opening angle, and other relevant parameters. During the simulation, the module determines its velocity vector based on the glider's velocity and direction values, updating the moving distance and current position. Moreover, the module periodically checks whether the current node's coordinates are outside the preset range. If the node moves out of the threshold, the module will recalculate the velocity vector based on the opening angle, ensuring that the node performs a zigzag movement within the depth threshold, as shown in Fig.~\ref{p3.5a}.

The motion principle of WGs is similar to UGs. As represented in Fig.~\ref{p3.5c}, WGs pick up the speed from waves and achieve movement on the sea. Then solar energy provides energy for WGs' various functions, such as data collection, communication, and localization. In Aqua-Sim~FG, we design {\tt{Aqua-Sim~FG::WG Mobility}} module to simulate WG movement trajectory. Users need to set the initial WG state, including velocity and movement direction. To simulate the WG movement state more realistically, we introduce a new wave model {\tt{Aqua-Sim~FG::WaveUtil}} in Aqua-Sim~FG. In this way, the movement module calculates a WG's current three-dimensional position based on the wave model and two-dimensional sea surface coordinates.   

AUVs in real UANs generally follow preset paths to perform underwater tasks, as shown in Fig.~\ref{p3.5b}. In Aqua-Sim~FG, the mobile trajectory is simulated by the {\tt{Aqua-Sim~FG::UV Mobility}} module. When the system starts, the UV-mobility module reads movement instructions from a text, which indicates specific movement modes and parameters. The movement modes include straight line and curve two modes. For the straight line mode, users need to set the mobile velocity, direction, and their duration time in the text. For the curve mode, users need to set the linear velocity, angular velocity, pitch angle, and their duration time. The UV-mobility module saves current instructions and the location of an AUV. When the node calls the module to obtain the latest location, the UV-mobility module calculates the current location based on the saved information and time difference. In addition, after finishing the current instruction, the module can read the next instruction automatically to update the movement mode. If all instructions are executed, the AUV puts on standby.

\begin{figure}[t]
	\centering	

	\subfloat[Fixed packet header structure in Aqua-Sim~TG \label{p3.4a}]{\includegraphics[width=6cm]{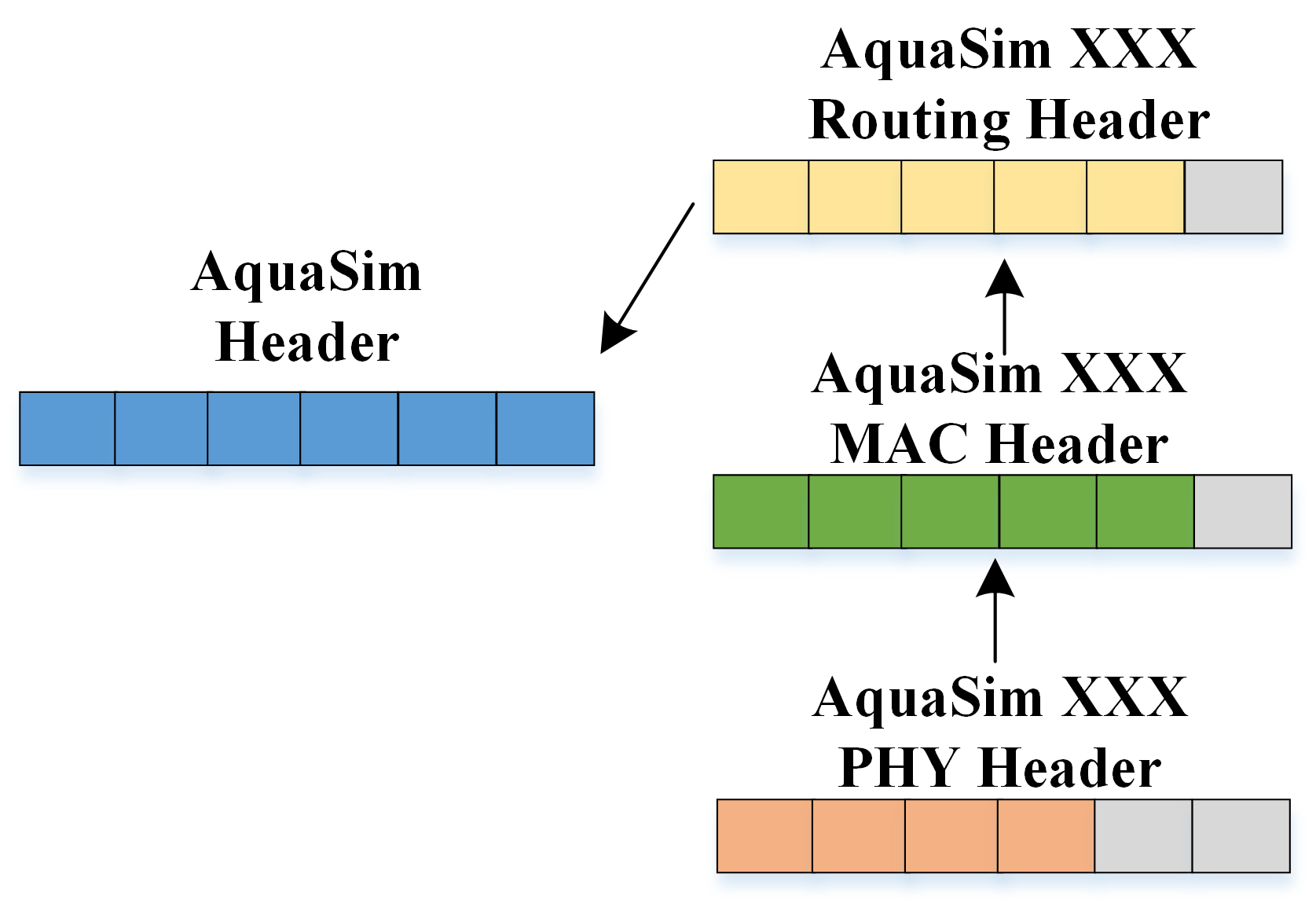}}	

	\subfloat[Adaptive packet header structure in Aqua-Sim~FG \label{p3.4b}	]{\includegraphics[width=8cm ]{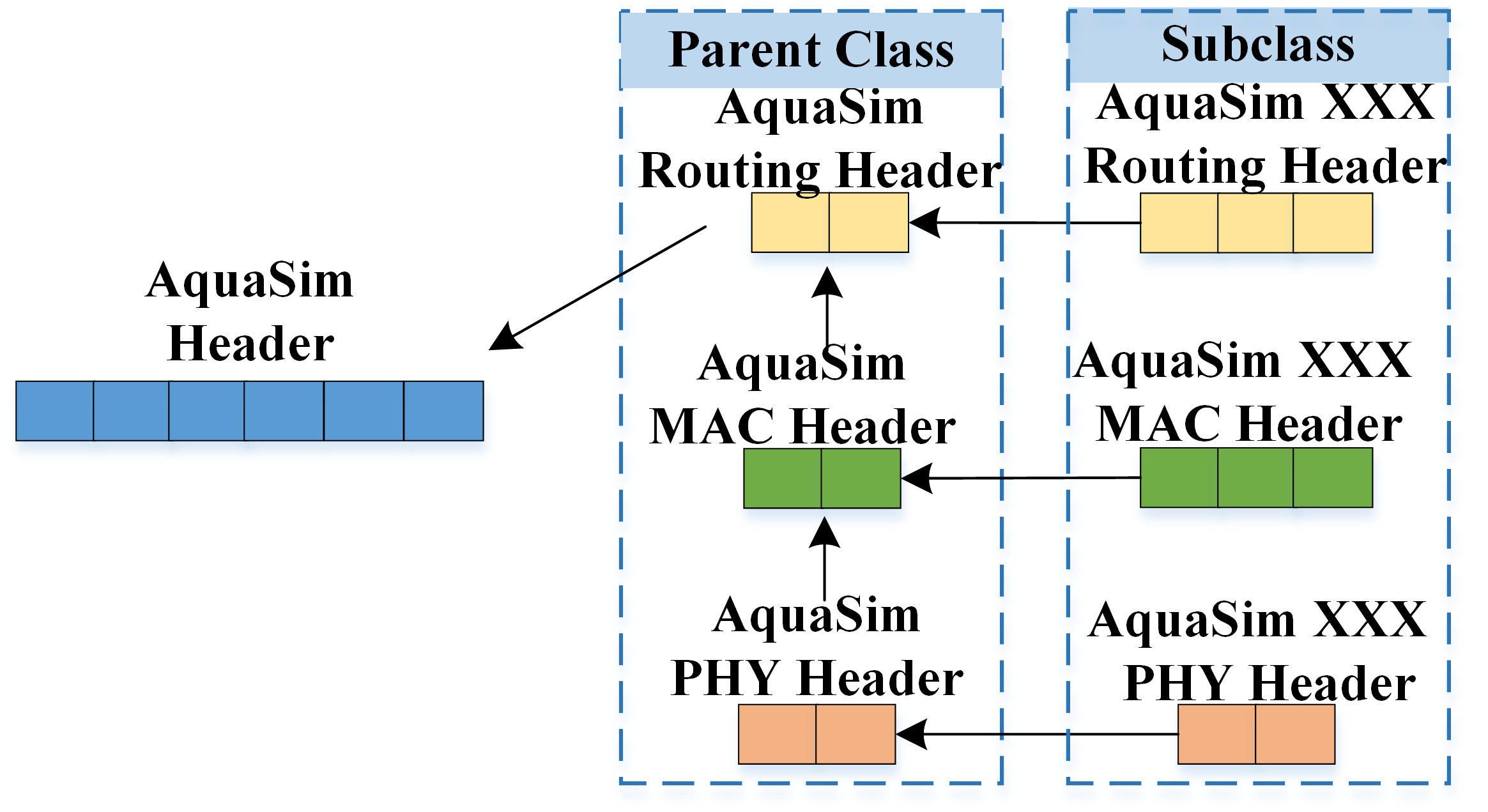}}				
	\caption{Comparison of the packet header structure between Aqua-Sim~TG and Aqua-Sim~FG. Gray blocks in Fig.~\ref{p3.4a} denote empty blocks without information. XXX denotes that protocols include various types of packets.}
	\label{p3.4}	
\end{figure} 

\subsubsection{Adaptive Packet Header Structure \label{sec::header}}
Previous generations of Aqua-Sim-series simulators adopt fixed packet header structures for different types of packets in each network layer, as shown in Fig.~\ref{p3.4a}. However, various packets require different header information in practice.

Take Vector-Based Forwarding routing protocol (VBF) \cite{r31} as an example: the header of interest packets should include the sink and sender location information to detect the forwarding pipe area; the header of data packets should include positions of the sender, the target, and the forwarder to determine the holding time. To ensure the normal operation of various packets in the current protocol, the previous Aqua-Sim~TG designs a long structure based on the header length sum of all packet types. Such a design introduces surplus space for packets with less header information, resulting in longer transmission delay.

In real UANs, although various packets employ different header information to achieve communication, some information is indispensable to all packets (called public information). For example, all packets should record the sender ID and sink ID in the header. Along this line, Aqua-Sim~FG distinguishes public header and private header to configure header information flexibly, as represented in Fig.~\ref{p3.4b}. The public header in each layer records the public information and is implemented in the parent class. The private header is implemented in the concrete packet's subclass and only records related header information specific to the current packet. In this way, we keep the public header's information unchanged and adjust the private header's information based on the current packet type. Compared with Aqua-Sim~TG, Aqua-Sim~FG provides a variable-length packet header, adapting to different types of packets.

\subsubsection{Cross-layer Information Interaction}

\begin{figure}[!t]
	\centering
	\includegraphics[width=9cm]{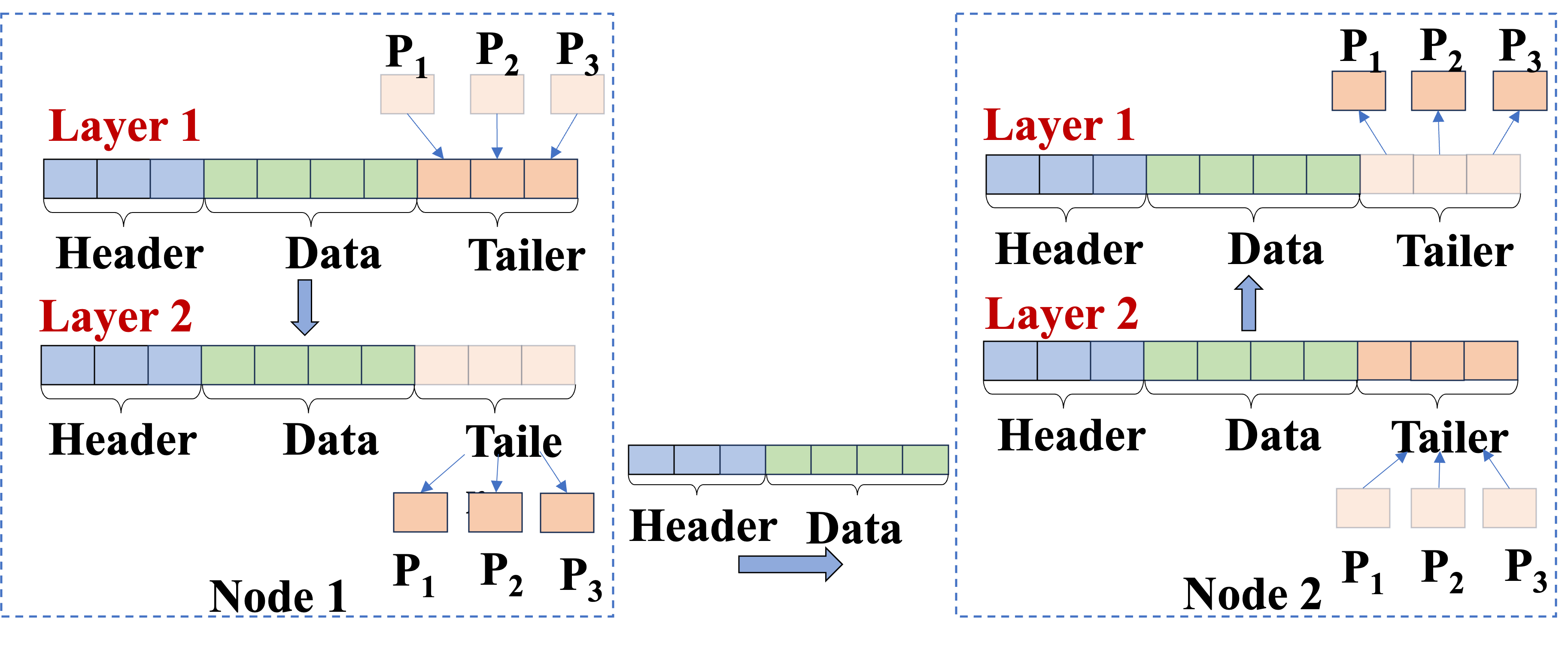}
	\caption{An example of cross-layer parameter passing via tailer structure in Aqua-Sim~FG. $P_n$ denotes the $n$-th parameter. Tailer information does not pass across nodes.}
	\label{p3.6}
\end{figure}

In recent years, researchers propose various cross-layer protocols to improve UANs performance with different technologies' advantages \cite{r17}, \cite{r18}. Such cross-layer protocols generally require exchanging information between layers, but interaction layers and parameters are all different, increasing the protocol implementation complexity.   
To avoid the complex parameter passing among functions, Aqua-Sim~FG introduces a new structure {\tt{Aqua-Sim~FG::Tailer}} into packets as shown in Fig. \ref{p3.6}. Users can put parameters that need to be passed across layers into the tailer and process the tailer like the header. In this way, different layers can extract the required information from the packets without modifying the program from the function level. It is worth noting that the tailer information is passed only across layers but not across nodes, thus it does not increase extra packet transmission load.        

\subsection{New Communication Features} 

\subsubsection{Subcarrier-level Frequency Spectrum Configuration}
Previous generations of simulators provide channel-level division schemes, which means the whole bandwidth is divided into specific several sub-channels, and users allocate these sub-channels for different nodes. However, the existing frequency allocation scheme is gradually toward to sub-carrier level \cite{r25}, \cite{r26}, and the channel-level simulation fails to simulate these novel schemes.

Compared with previous simulators, which only configure the number and bandwidth of sub-channels, Aqua-Sim~FG provides more detailed frequency resource configuration. It supports configuring frequency resources from the sub-carrier level, including the whole bandwidth, the number of sub-carriers, sub-carrier interval, and sub-carrier guard time. Such a fine-grained configuration can reflect the effects of frequency-related resources on transmission rate, delay, and energy consumption. During the specific transmission process, nodes configure the final sub-carrier level frequency spectrum information based on upper-layer protocols, input the information into the physical layer's header, and then send the packet to the receiver via the channel module. Based on the header's information, the receiver evaluates whether the current frequency spectrum has been occupied by other nodes and further estimates whether packet collision occurs. In this way, Aqua-Sim~FG simulates more complex schemes and assists in exploring the relationship between frequency spectrum resources and transmission performance from various fine-grained perspectives.

\subsubsection{Multiple Modulation Modes}

To deal with the spatial and temporal variation of underwater acoustic channels, adaptive modulation techniques are proposed for UANs \cite{r27}, \cite{r28}, which vary transmission parameters based on channel conditions, improving communication efficiency. Moreover, the technique has been applied to actual underwater acoustic modems \cite{r29}.   

In Aqua-Sim~FG, we design a new mode configuration function to expand simulation ability. Specifically, we consider five frequently-used modulation modes, including BPSK, QPSK, 8QAM, 16QAM, and 64QAM. The code rate of the above modes increases successively; the larger code rate means a larger transmission rate, but modes providing a larger transmission rate require higher channel quality. Users can configure different modulation modes for packets in Aqua-Sim~FG, and the mode information is recorded in the packet header. When the packet arrives at the physical layer, the sender simulates the transmission rate based on the mode information and sends the packet to the receiver. Then, the receiver's physical layer calculates the received Signal to Noise Ratio (SNR) threshold based on the mode information. After that, the receiver compares the SNR threshold with the actual receiving SNR or calculates Bit Error Rate (BER) to determine whether the packet can be received (the packet can be received only when the actual SNR or BER is larger than the threshold). In this way, Aqua-Sim~FG can simulate adaptive modulation schemes and evaluate their performance.                  
\subsubsection{Multiple Propagation Models}

Aqua-Sim~FG provides multiple propagation models, including range-based propagation model, Thorp propagation model, and Bellhop propagation model, to simulate underwater acoustic signal propagation with different degrees. Users can configure the specific propagation models based on their own simulation requirements.

{\tt{Aqua-Sim~FG::RangePropagtion}} as the range-based propagation model simulates the signal propagation most easily: users can preset a propagation range threshold, if the distance between two nodes is larger than the range threshold, these two nodes fail to receive signals from each other. Under such a condition, distance is the unique factor that affects the success of communication.      

{\tt{Aqua-Sim~FG::ThorpPropagtion}} adopts Thorp model \cite{r30} to simulate the signal propagation, and the attenuation is calculated as follows:

\begin{equation}
    A(l,f)=A_0d^ka(f),
    \label{eq1}
\end{equation}
where $d$ is the propagation distance and $f$ is the sending frequency. $A_0$ is a constant, $k$ is the attenuation factor, and $a(f)$ denotes the absorption factor. By employing Thorp model, this model considers both distance, sending frequency, and partial environmental correlation factors to evaluate the signal attenuation. Based on the attenuation, the receiver further calculates SNR or BER to determine whether the packet can be received. Compared with the range-based propagation model, Thorp model provides a more detailed propagation simulation.

{\tt{Aqua-Sim~FG::BellhopPropagtion}} integrates Bellhop and simulates signal propagation based on real environmental information. Bellhop is a classical underwater acoustic toolbox, which can simulate the number of propagation paths, the incidence angle, the amplitude, and the arrival delay based on environmental information (such as sound speed profile, submarine topography, and absorption coefficient). In this way, Bellhop calculates sound field information in a heterogeneous environment, including transmission loss, impulse response, and eigenray propagation path. Therefore, Bellhop propagation model simulates signal propagation much closer to real ocean scenarios. {\tt{Aqua-Sim~FG::Bellhop3DPropagtion}} integrates Bellhop3D and provides three-dimensional sound tracking. Both the above two models provide transmission loss for the physical layer to calculate SNR or BER and assist the physical layer in determining whether the packet can be received. 

\section{Simulation Results}

In this section, we first compare Aqua-Sim~FG with real field experiments to evaluate its authenticity. Then, we compare Aqua-Sim~FG with other simulation schemes to analyze Aqua-Sim~FG's advantages.   

\subsection{Analysis of Aqua-Sim~FG Authenticity}

\begin{figure}[t]
	\centering	

	\subfloat[Test equipment \label{p5.1a}]{\includegraphics[width=3.5cm]{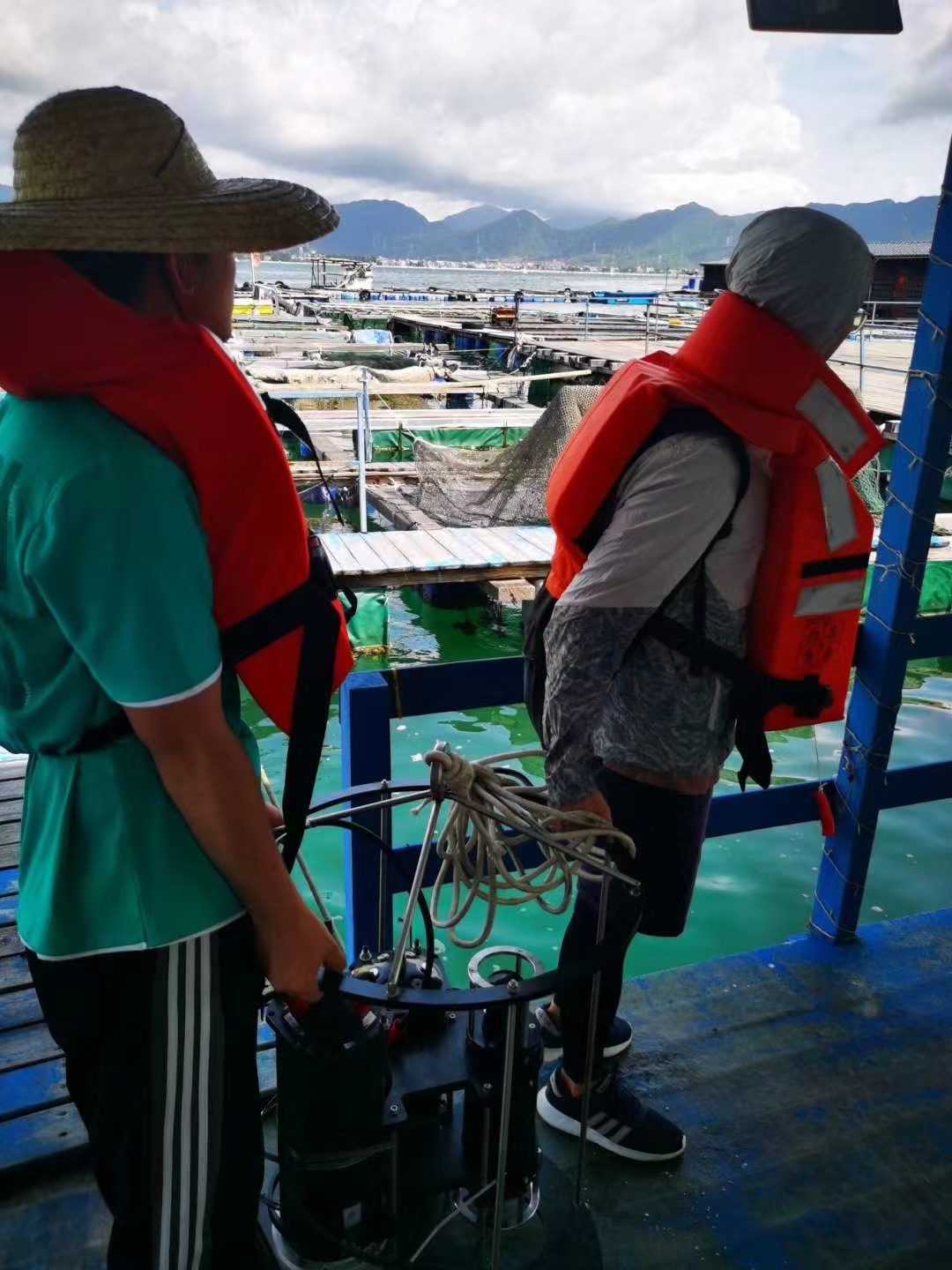}}	
	\hspace{0.3cm}	
	\subfloat[Deployment of the experiment \label{p5.1b}	]{\includegraphics[width=3.4cm,height=4.65cm]{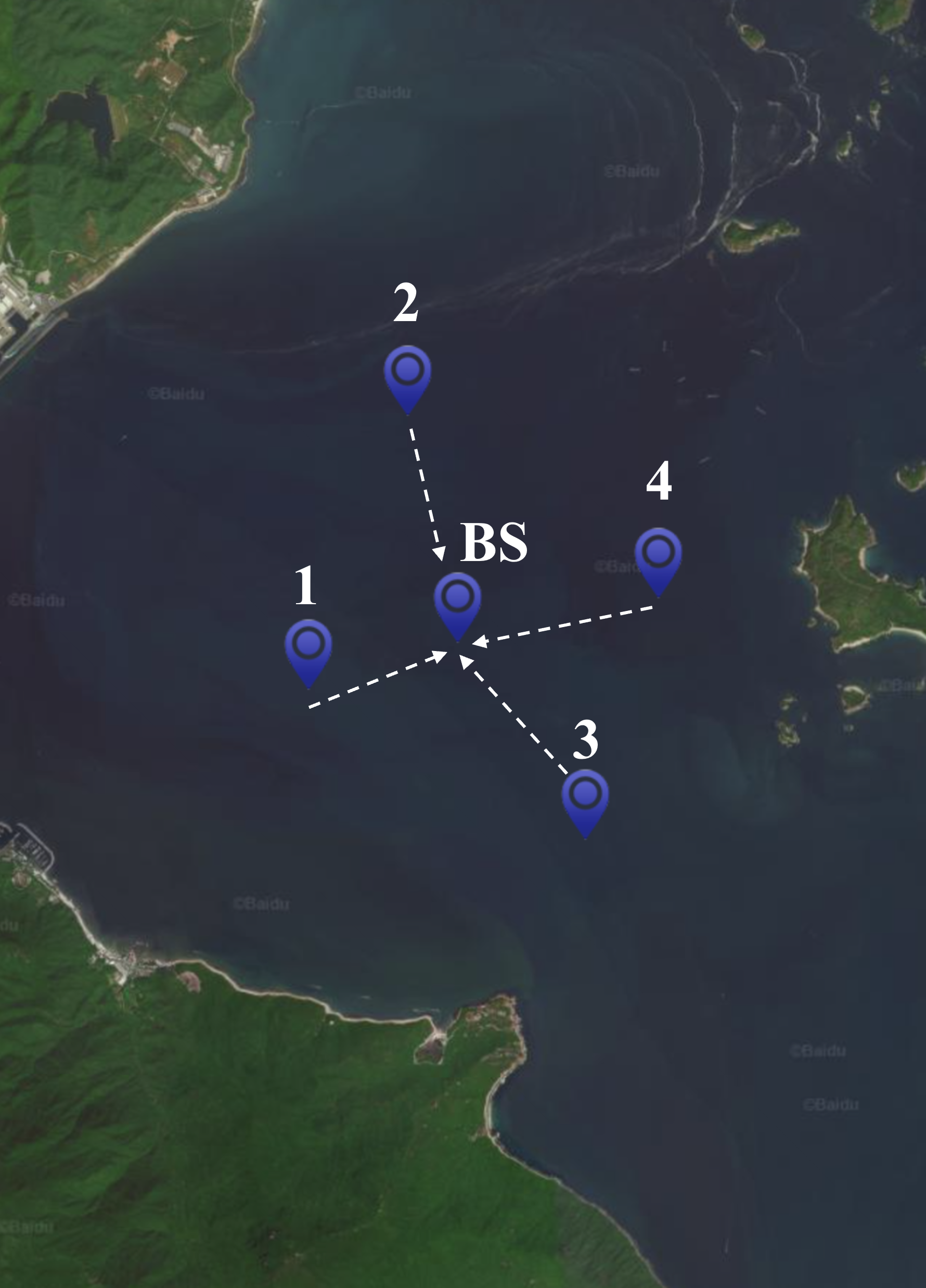}}		

 	\subfloat[Throughput comparison \label{p5.1c}]{\includegraphics[width=4.3cm]{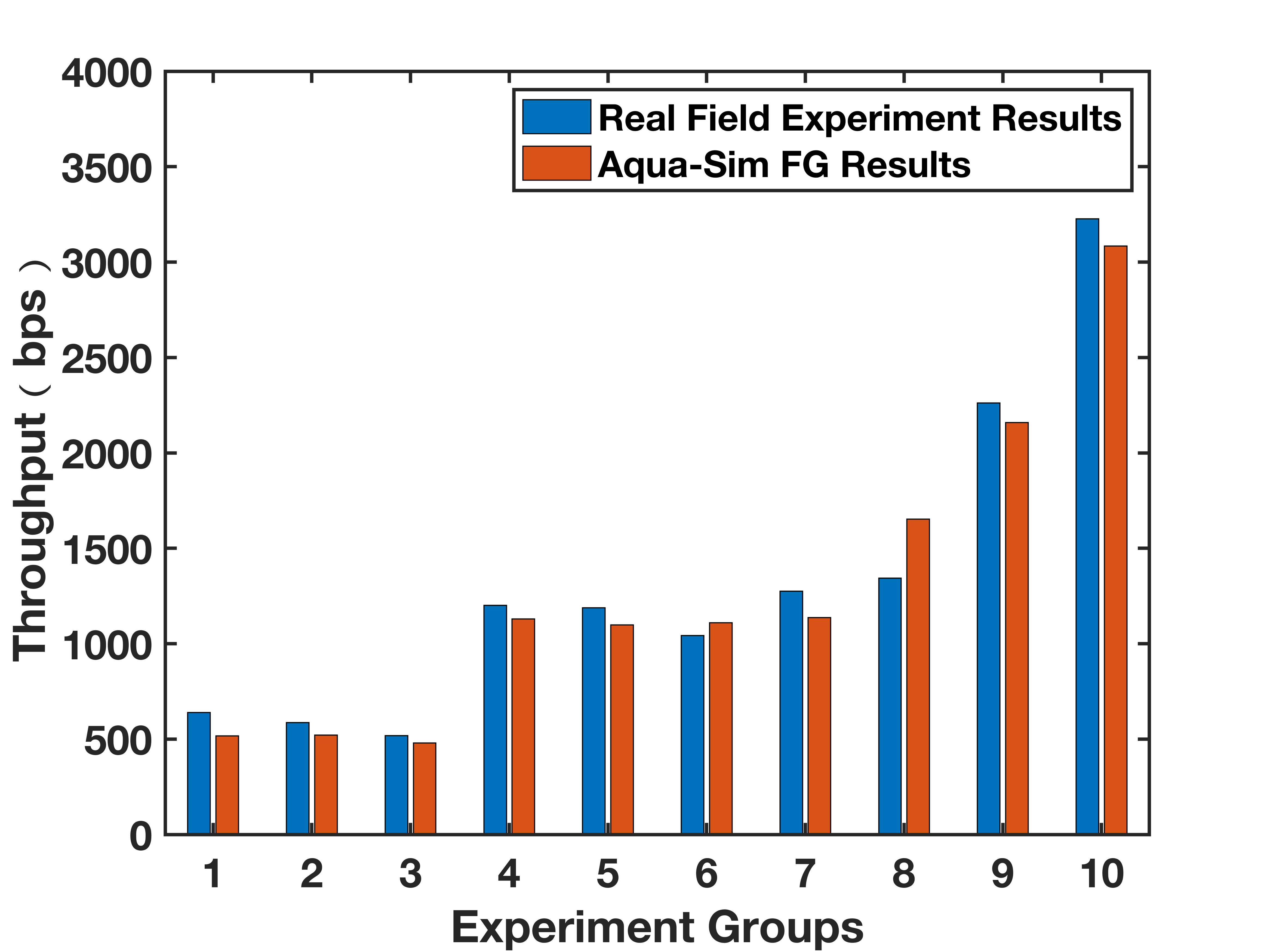}}	
	\hspace{0.1cm}	
	\subfloat[End-to-end delay comparison \label{p5.1d}	]{\includegraphics[width=4.3cm ]{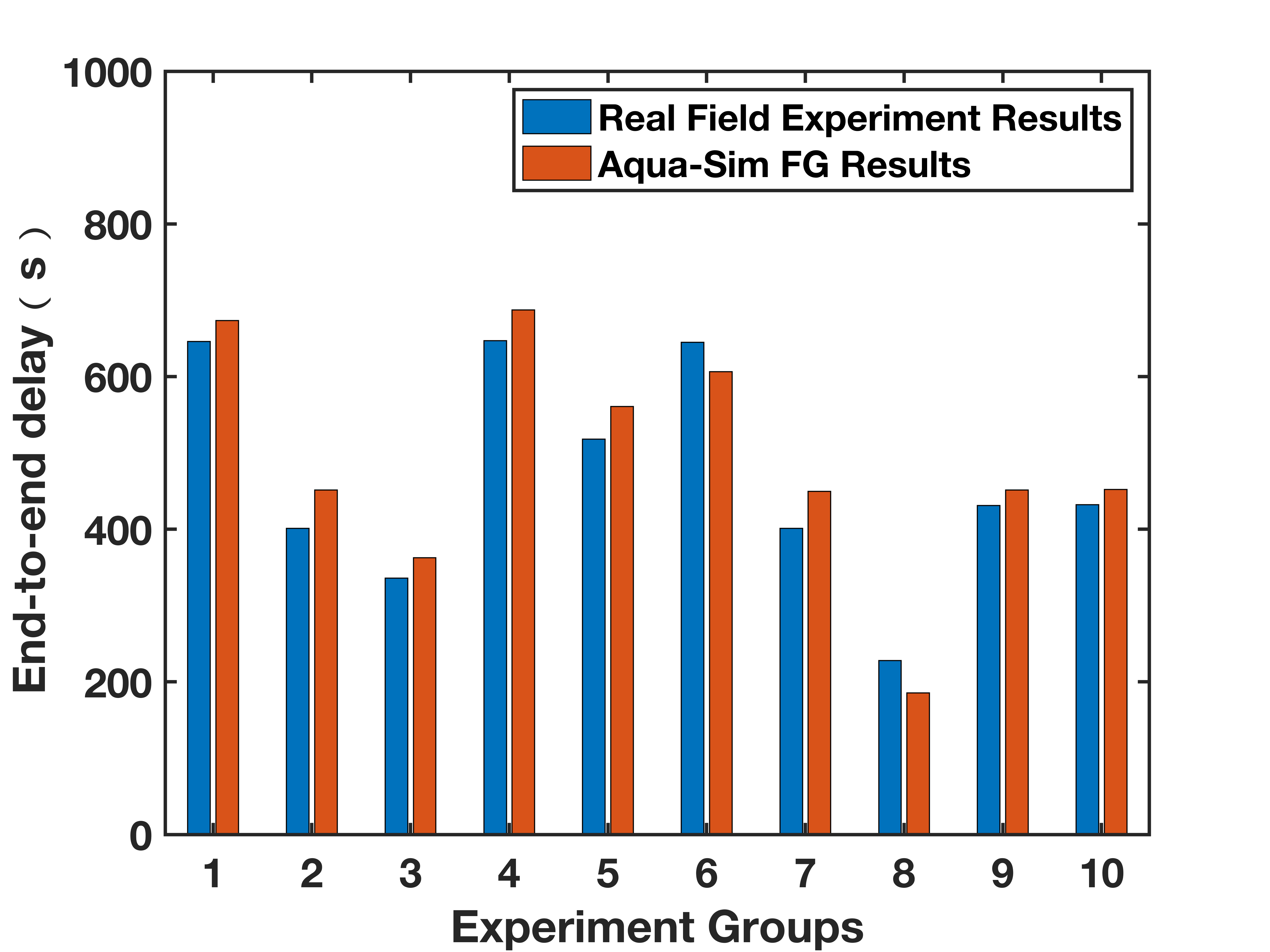}}
	\caption{Comparison 1 of real field experiments and Aqua-Sim~FG. The real network is a five-node cluster network deployed in Daya Bay, Guangdong, China, 2019. All nodes transmit packets to the base station. }
	\label{p5.1}	
\end{figure}

\begin{figure*}[t]
	\centering	

	\subfloat[Test equipment \label{p5.2a}]{\includegraphics[width=6cm]{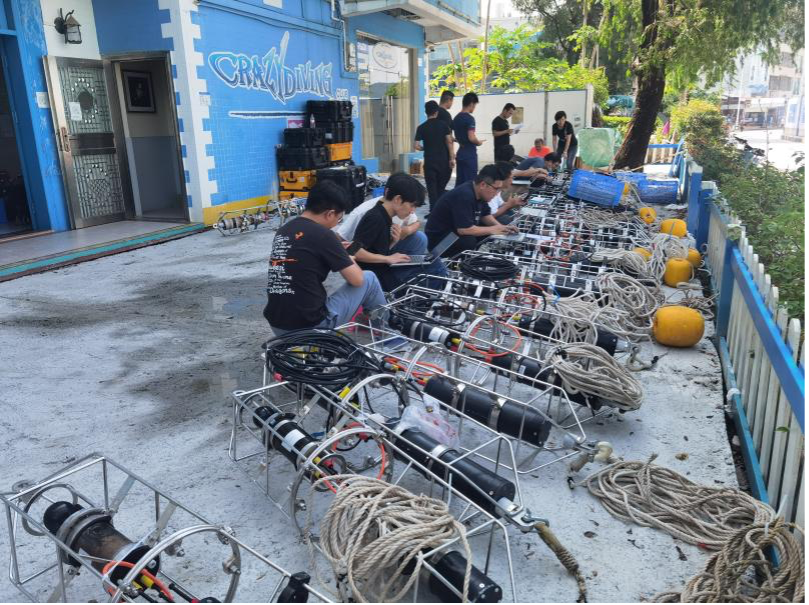}}	
	\hspace{0.1cm}	
	\subfloat[Deployment of the experiment \label{p5.2b}]{\includegraphics[width=5.02cm ]{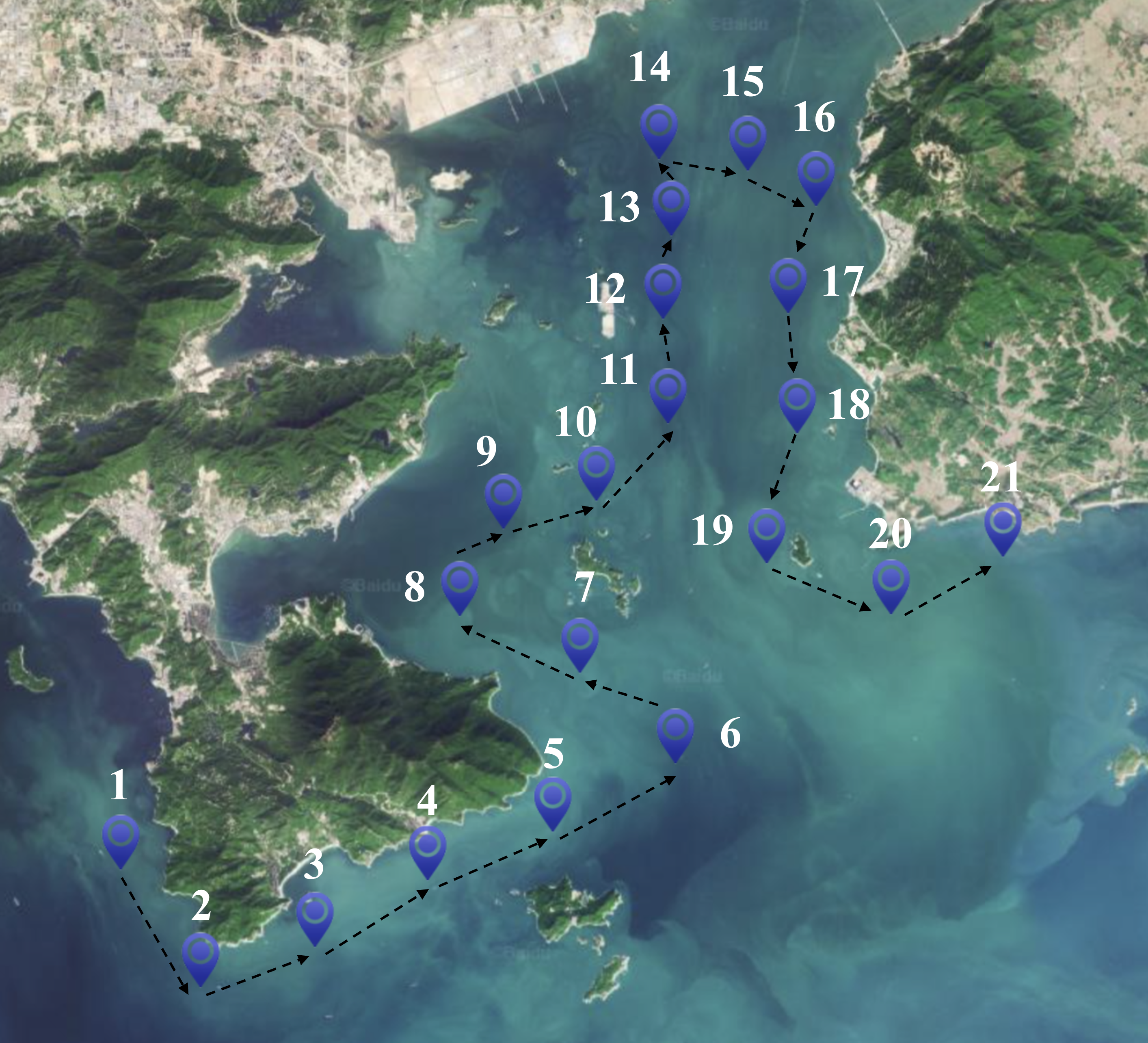}}	
 	\hspace{0.1cm}
 	\subfloat[Throughput comparison \label{p5.2c}]{\includegraphics[width=5.9cm]{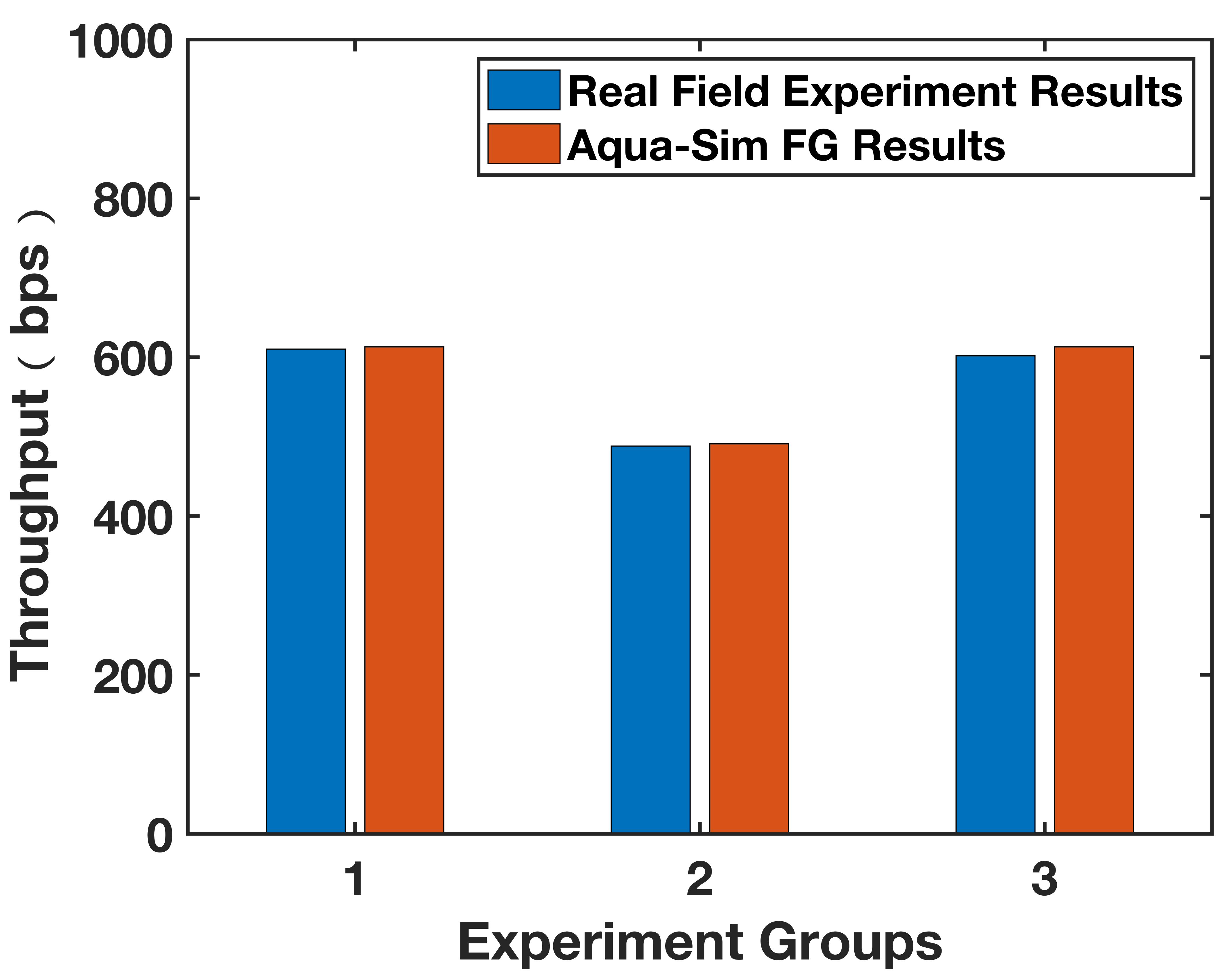}}	
	\caption{Comparison 2 of real field experiments and Aqua-Sim~FG. The real network is a 21-node string network deployed in Daya Bay, Guangdong, China, 2022. Packets are transmitted from node 1 to node 21.}
	\label{p5.2}	
\end{figure*}

We compare Aqua-Sim~FG with two real field experiments. As represented in Fig.~\ref{p5.1b}, the first real UAN is a five-node cluster network. Node 1 to node 4 are senders, transmitting packets to the Base Station (BS). The distance between each sender and BS is about 1 km. In Fig.~\ref{p5.2b}, the second real UAN is a 21-node string network. Packets are transmitted hop-by-hop from node 1 to node 21, and the average distance between two adjacent nodes is about 4 km. We simulate multiple groups of experimental results, each of which adopts different modulation modes (including BPSK, QPSK, 8QAM, 16QAM, and 64QAM) and sends different sizes of packets. As represented in Fig.~\ref{p5.1c}, Fig.~\ref{p5.1d}, and Fig.~\ref{p5.2c}, we can observe that Aqua-Sim~FG's results are relatively similar to real field experiments' results. In addition, these comparison results also demonstrate that Aqua-Sim~FG can configure different modulation modes and simulate UANs with different topologies and network scales.  

\begin{figure*}[t]
	\centering	

	\subfloat[Transmission loss of different propagation models \label{p5.3a}]{\includegraphics[width=5.5cm]{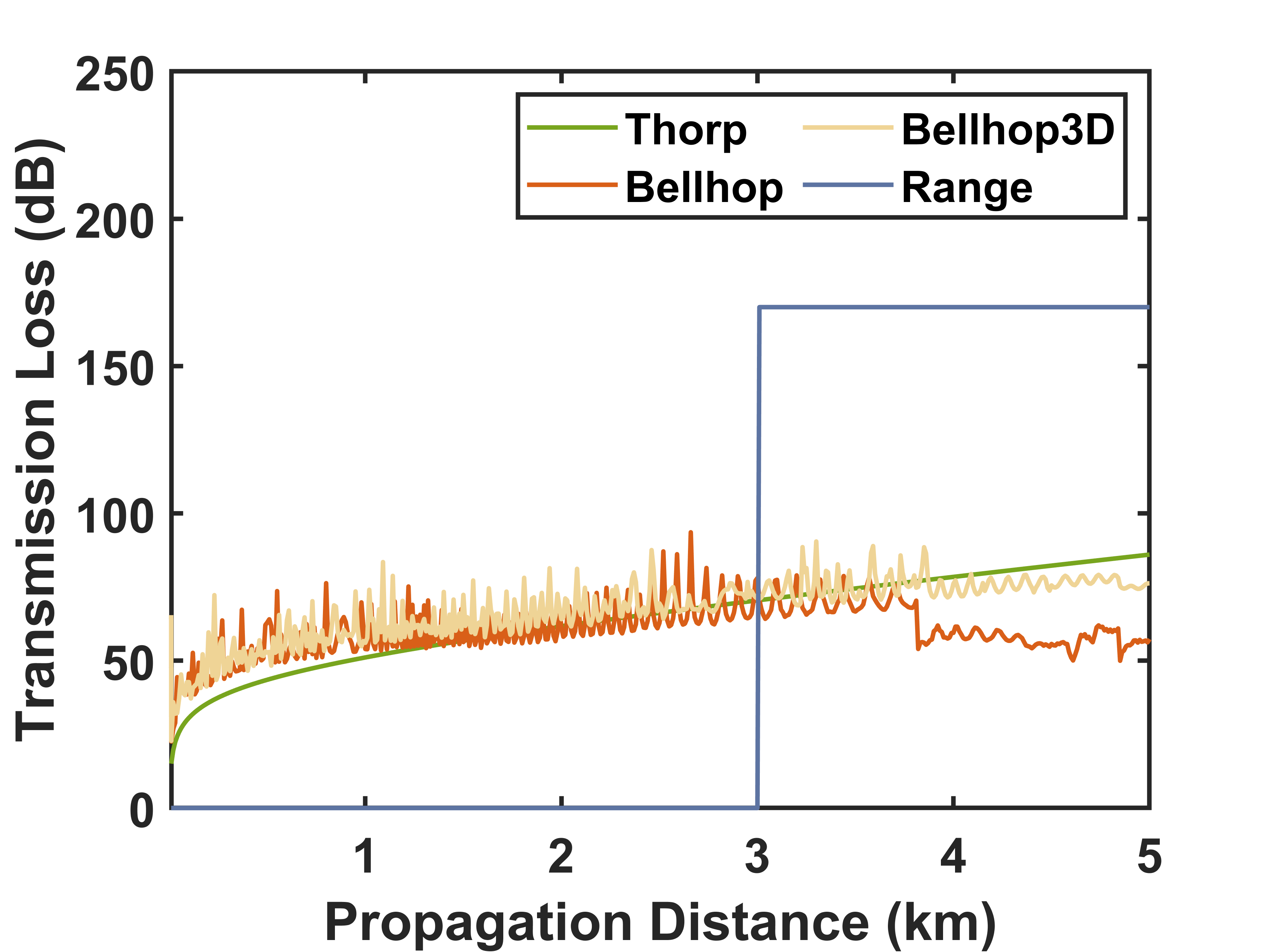}}	
	\hspace{0.1cm}	
	\subfloat[BER of different communication technologies \label{p5.3b}]{\includegraphics[width=5.5cm]{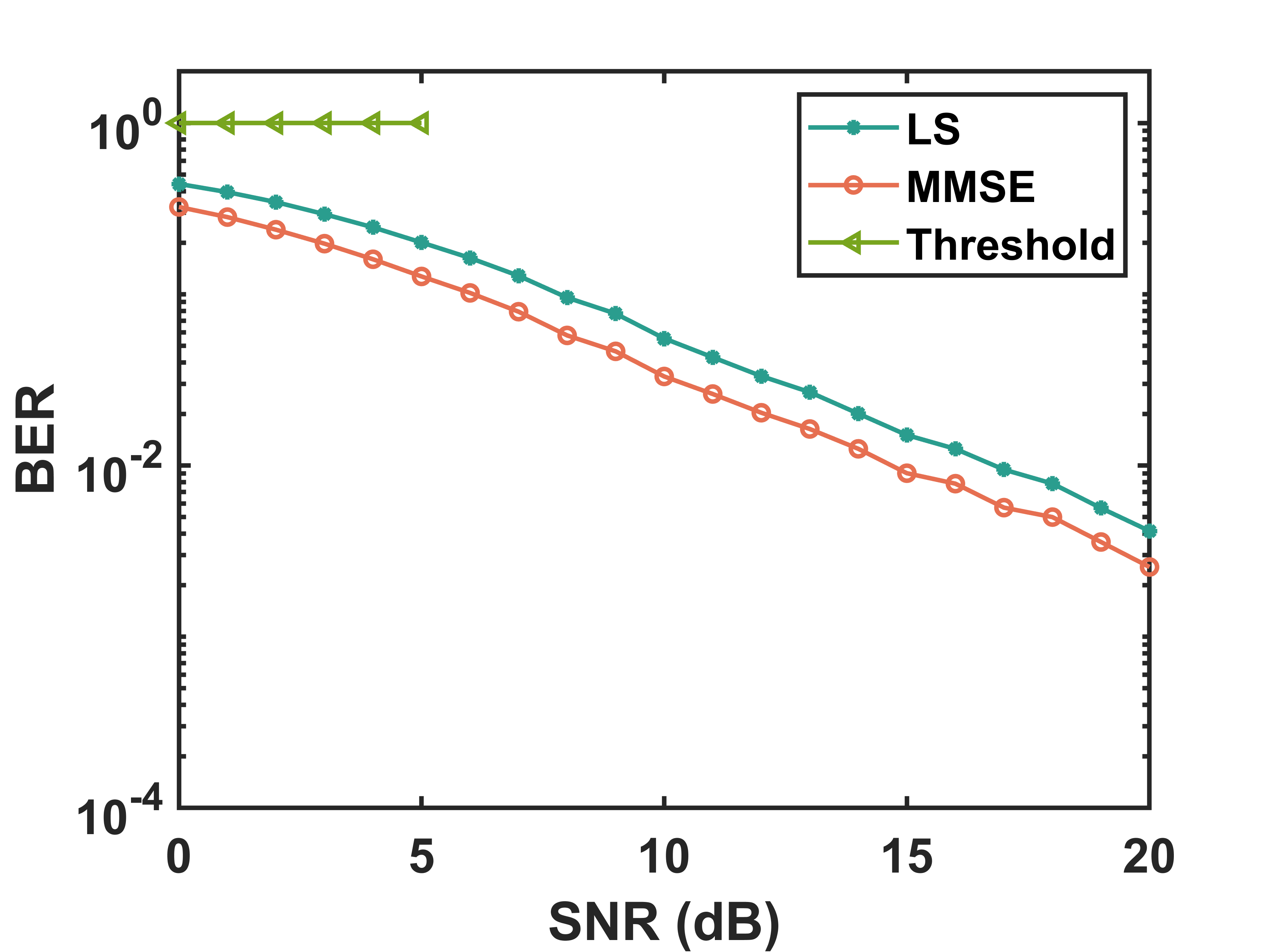}}	
    \hspace{0.1cm}
 	\subfloat[Throughput under different protocol and communication technology combinations \label{p5.3c}]{\includegraphics[width=5.5cm]{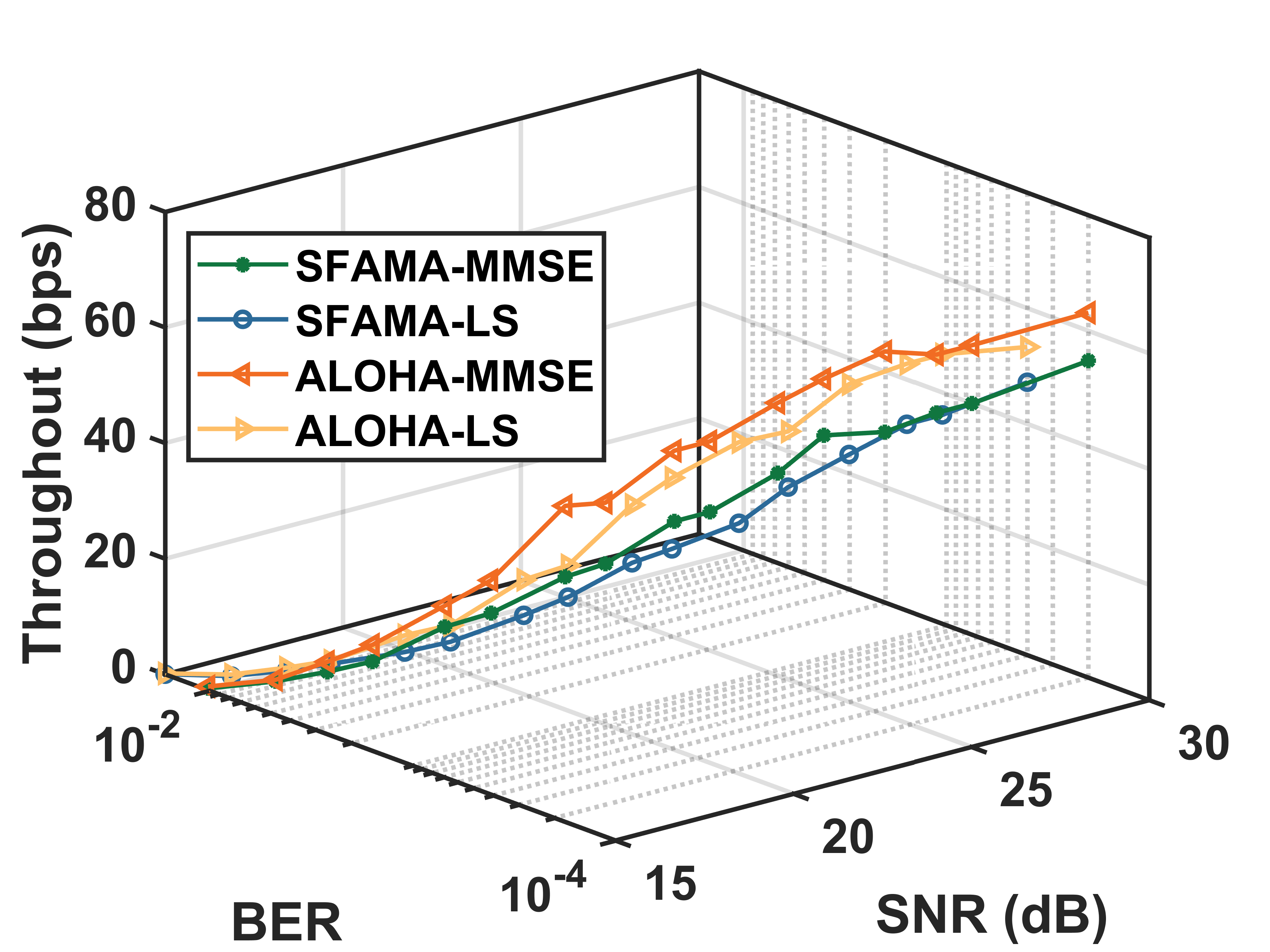}}	

	\caption{Interaction effects between communication technologies and network protocols on performance simulation results.}
	\label{p5.3}	
\end{figure*}

\subsection{Interaction Effects between Communication and Network}

In this set of experiments, we discuss the effects of propagation models on channel quality simulation. Then, we demonstrate the necessity of channel quality for communication technology simulation. Based on specific BER provided by communication technology, we further evaluate the performance of network protocols.   

As represented in Fig.~\ref{p5.3a}, we employ transmission loss (TL) to evaluate channel quality. The larger TL means the low channel quality. Range model only considers propagation distance to determine whether the packet can be received. We set 3 km as the threshold, when the distance is less than the threshold, the model simulates that the channel has no loss. On the contrary, the model simulates that the signal loses all sound source energy (around 170~dB), indicating no packets can be received. TL values of Thorp model increase as the propagation distance becomes longer, which conforms to the rule that signal energy decays continuously as propagation distance increases. However, in actual UANs, channel quality is affected by various factors, and the relationship between attenuation and distance is not absolute. Unfortunately, Thorp model only simply displays the variation tendency of the channel quality, failing to demonstrate more detailed information. Bellhop and Bellhop3D models simulate channel quality based on realistic marine data for a region of the South Pacific \cite{r32}. We set the sender and the receiver at a depth of 50~m. Compared with other models, these two models provide more detailed channel qualities. Moreover, 3D model considers more terrain factors and three-dimensional sound speed information than 2D model. Therefore, Bellhop3D model results differ from Bellhop model results. These propagation models provide various degrees of TL, and users can select the corresponding model based on their requirements, further calculating SNR to achieve channel simulation.

Based on different channel qualities provided by various propagation models, we further compare BER calculated by different communication technologies. The easiest method is to set an SNR threshold as shown in Fig.~\ref{p5.3b} (the threshold is set to 5~dB in this experiment). If the SNR value is smaller than 5~dB, the BER is set to 1. On the contrary, the BER is set to 0. Least Squares (LS) and Minimum Mean Square Error (MMSE) are two classical channel estimation methods \cite{r33}. Compared with the threshold method, LS and MMSE employ pilot symbols to estimate the channel and calculate BER, providing more reasonable information. However, due to the different accuracy of these communication technologies, they calculate different BER values. 

As represented in Fig.~\ref{p5.3c}, we simulate UAN throughput of Aloha and Slotted Floor Acquisition Multiple Access (SFAMA) \cite{r34} MAC protocols and static routing based on different channel quality and BER. The network is a 5-node cluster UAN, in which node 1 to node 4 are senders, and node 5 is the receiver. The packet size is 100~B, the transmission rate is 1500 bps, and the propagation distance is around 1~km. Under such a condition, although differences in protocols introduce different throughput results, channel quality and final BER also affect the throughput of an individual protocol. This is because channel quality and BER determine packet error, which affects retransmission number, delay, and queue congestion of packets, finally influencing throughput results.

All the above results demonstrate that different combinations of channel quality, communication technology, and protocol affect simulation accuracy. An individual simulation of each technology fails to reflect real UAN performance. In this generation simulator, Aqua-Sim~FG provides various propagation models to simulate underwater acoustic channels with different degrees. Based on a specific channel environment, Aqua-Sim~FG can simulate communication technologies and network protocols integrally, which assists users in analyzing communication technology and network protocols' interaction effects, and further improving troubleshooting and optimization ability. 
        
\subsection{Performance of AI Methods in UANs}

In this set of experiments, we select the algorithm (DCMD) proposed in \cite{r12} as an example of underwater AI methods. DCMD is a DRL-based data collection algorithm for UANs, in which AUVs constantly exchange state information and collaboratively determine collection trajectories to achieve a high collection rate with low costs. Under such a scenario, we implement DCMD with three AUVs and 25 sensor nodes in Aqua-Sim~FG and Python respectively, and focus on information exchange and training convergence to evaluate the effects of UANs' characteristics on DCMD's performance. 

\begin{figure}[t]
	\centering	

	\subfloat[Effects of UAN characteristics on observations \label{p5.4a}]{\includegraphics[width=4.2cm]{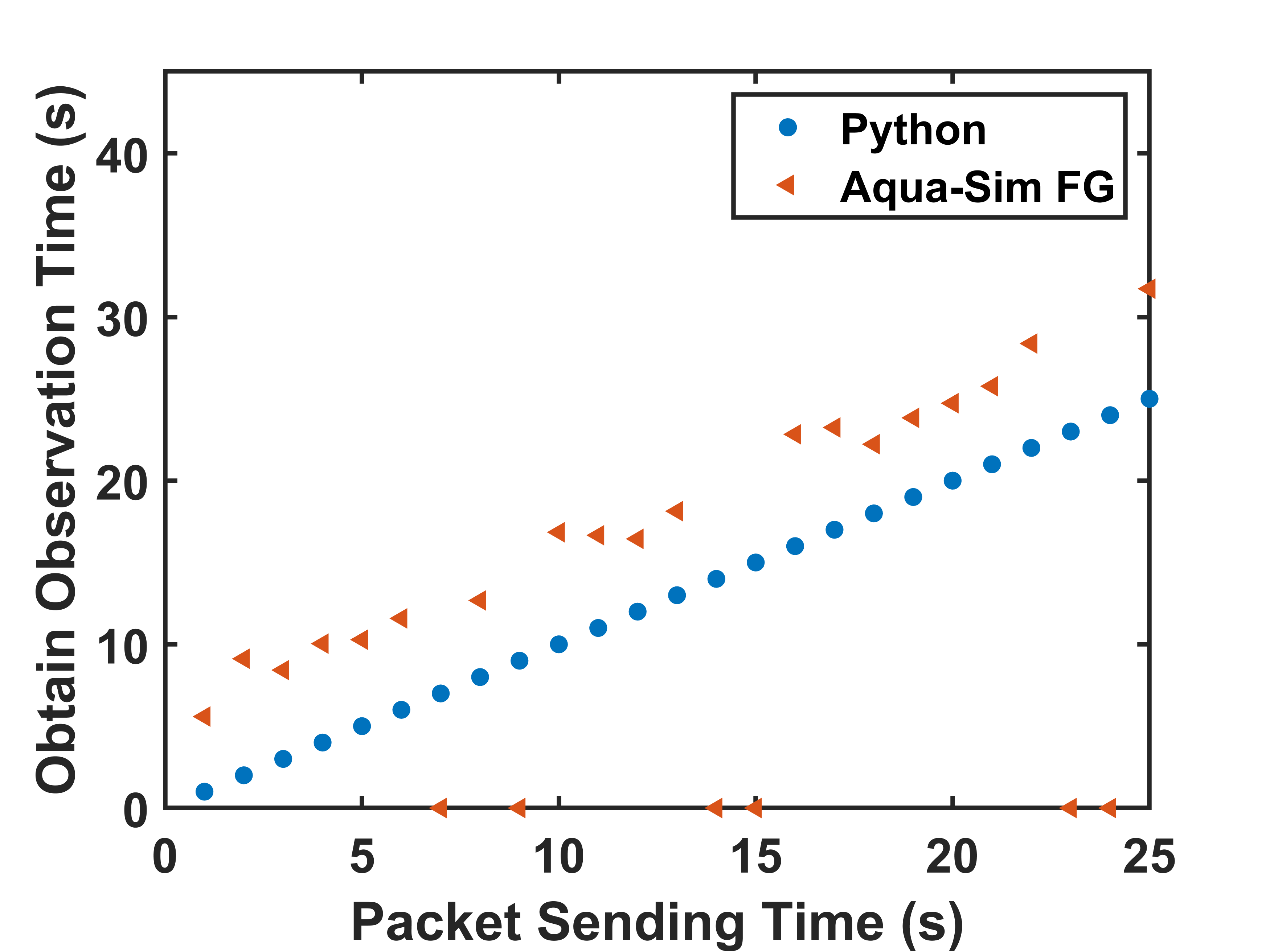}}	
	\hspace{0.1cm}	
	\subfloat[Training Results \label{p5.4b}]{\includegraphics[width=4.2cm]{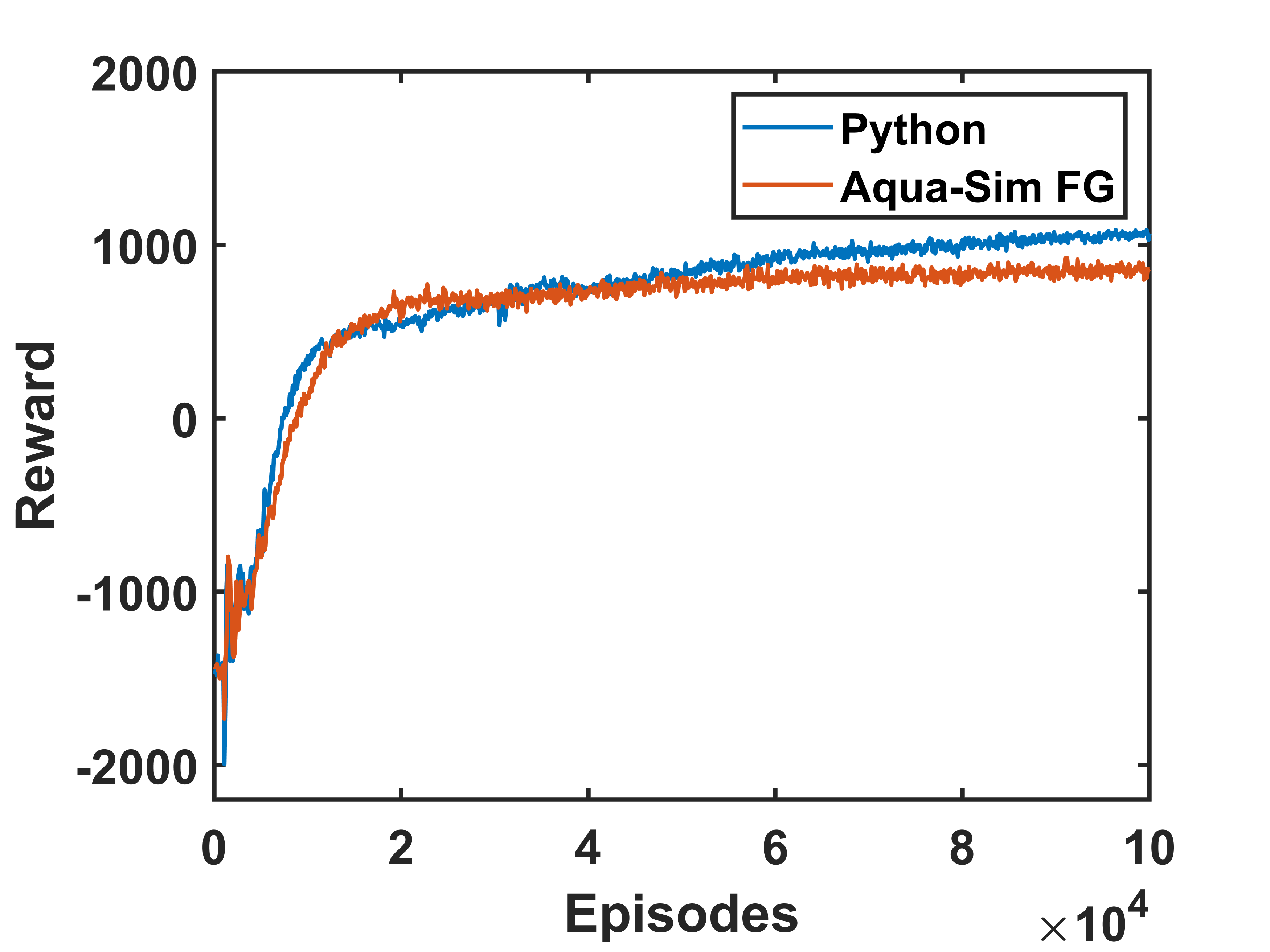}}	
  
	\caption{AI method implementation results’ comparison of Python and Aqua-Sim~FG. }
	\label{p5.4}	
\end{figure}

We first simulate the scenario in which AUVs exchange information to obtain observations. \ref{p5.4a} shows different results of Python and Aqua-Sim~FG. During the implementation, Python ignores the long propagation delay of underwater acoustic channels and assumes that an AUV can obtain the observation (information from other AUVs) immediately after sending the exchange demand, which is not in line with the actual UAN situation. Compared with Python, Aqua-Sim~FG can simulate UAN characteristics more authentically. For Aqua-Sim~FG's some markers, we can observe the y-axis value is larger than the x-axis value, which indicates an AUV waits for a long time to obtain observation information after sending the requirement packet. Moreover, some markers have no y-axis value, which means Aqua-Sim~FG also simulates packet loss of UANs for AI methods. Under such a condition, we further compare their training results.

Based on different information exchange results shown in Fig.~\ref{p5.4a}, Python and Aqua-Sim~FG further provide different training results in Fig.~\ref{p5.4b}. We can observe that Aqua-Sim~FG enters the convergence state earlier than Python, and its reward results are also smaller. This is because Aqua-Sim~FG simulates UANs' long propagation delay and packet loss, which introduces delayed and missing observations, reducing AI methods' performance. In this way, Aqua-Sim~FG provides more realistic UAN scenarios for AI methods than Python, assisting users in exploring more practical UAN AI problems. 
\begin{table*}[t]	
	\center
	\caption{Header size and transmission delay of various packets in different protocols.}
	\begin{threeparttable}
		\begin{tabular}{cccccccccccccccccccccccccccc}\toprule
			\multirow{2}{*}{}&\multicolumn{5}{c}{GOAL}&\multicolumn{5}{c}{SFAMA}&\multicolumn{4}{c}{VBF}\\
			\cmidrule(lr){2-6}\cmidrule(lr){7-11}\cmidrule(lr){12-15}
			&\makecell[c]REQ\tnote{3}&REP&DATA&ACK&TG\tnote{4}&RTS&CTS&DATA&ACK&TG&INTEREST&READY&DATA&TG
			\\ \midrule
			HS (B)\tnote{1}&53&28&20&9&92&8&8&7&7&12&31&19&43&79 \\ 
			TD (s)\tnote{2}&0.85&0.45&0.32&0.14&1.47&0.13&0.13&0.11&0.11&0.19&0.50&0.30&0.69&1.26\\ 
		\bottomrule	
		\end{tabular}
		\begin{tablenotes}
			\footnotesize
			\item[1] Header Size (HS) denotes the sum of public header size and subheader size for various packets.
			\item[2] Transmission Delay (TD) records the time spent to transmit header of different packets.
            \item [3] Due to the adaptive header structure, different types of packets in Aqua-Sim~FG have different header sizes. 
            \item[4] TG means Aqua-Sim~TG, in which each type of packet adopts a fixed header structure. All packets have the same header length. 
		\end{tablenotes}
	\end{threeparttable}
	
	\label{t1}	
\end{table*}

\subsection{Other Functions}

\begin{figure}[t]
	\centering	
    \includegraphics[width=8cm]{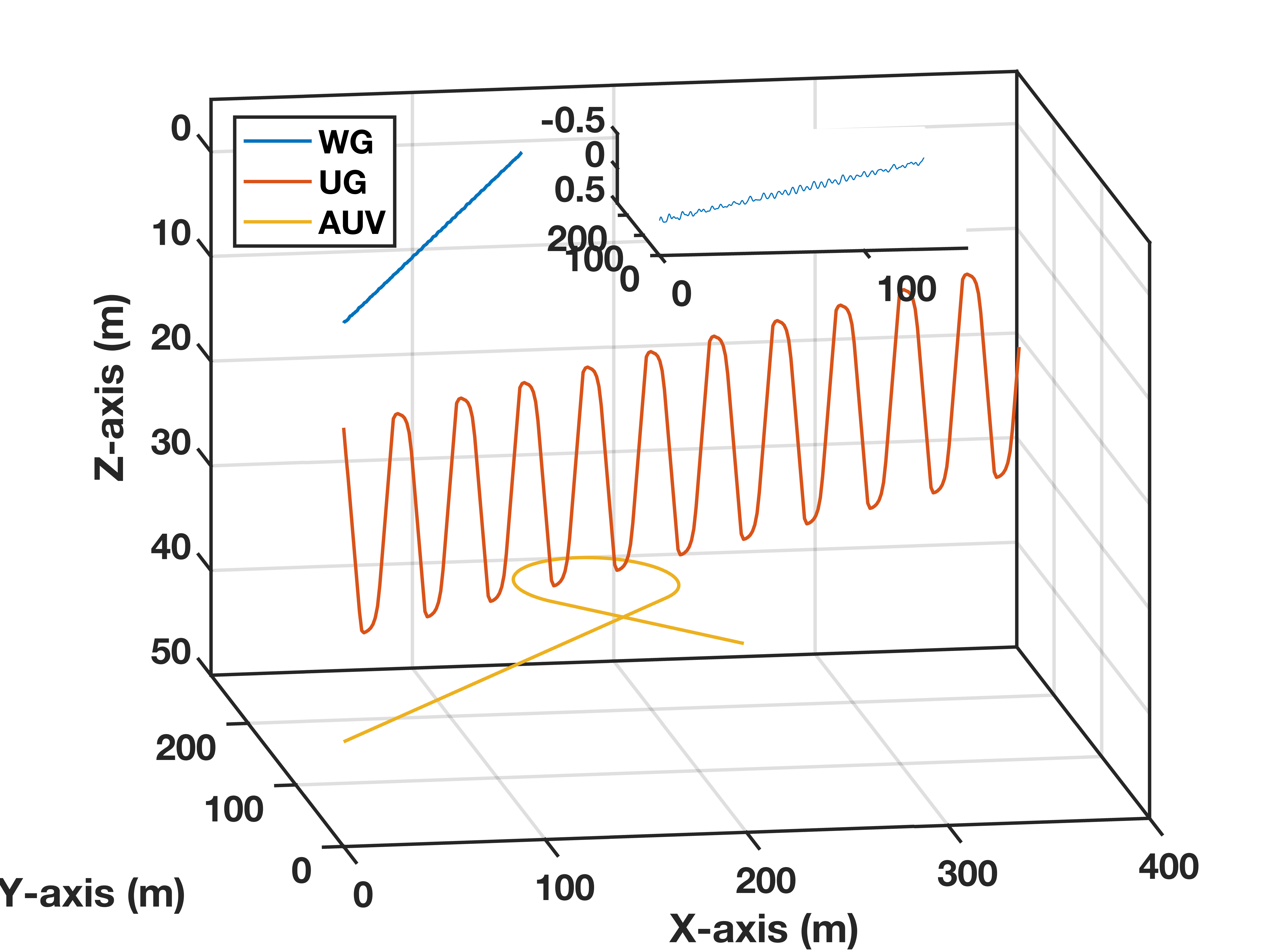}
	\caption{Movement trajectories of different underwater vehicles.}
	\label{p5.5}	
\end{figure}

\begin{figure}[t]
	\centering	

	\subfloat[Test equipment \label{p5.6a}]{\includegraphics[width=8cm]{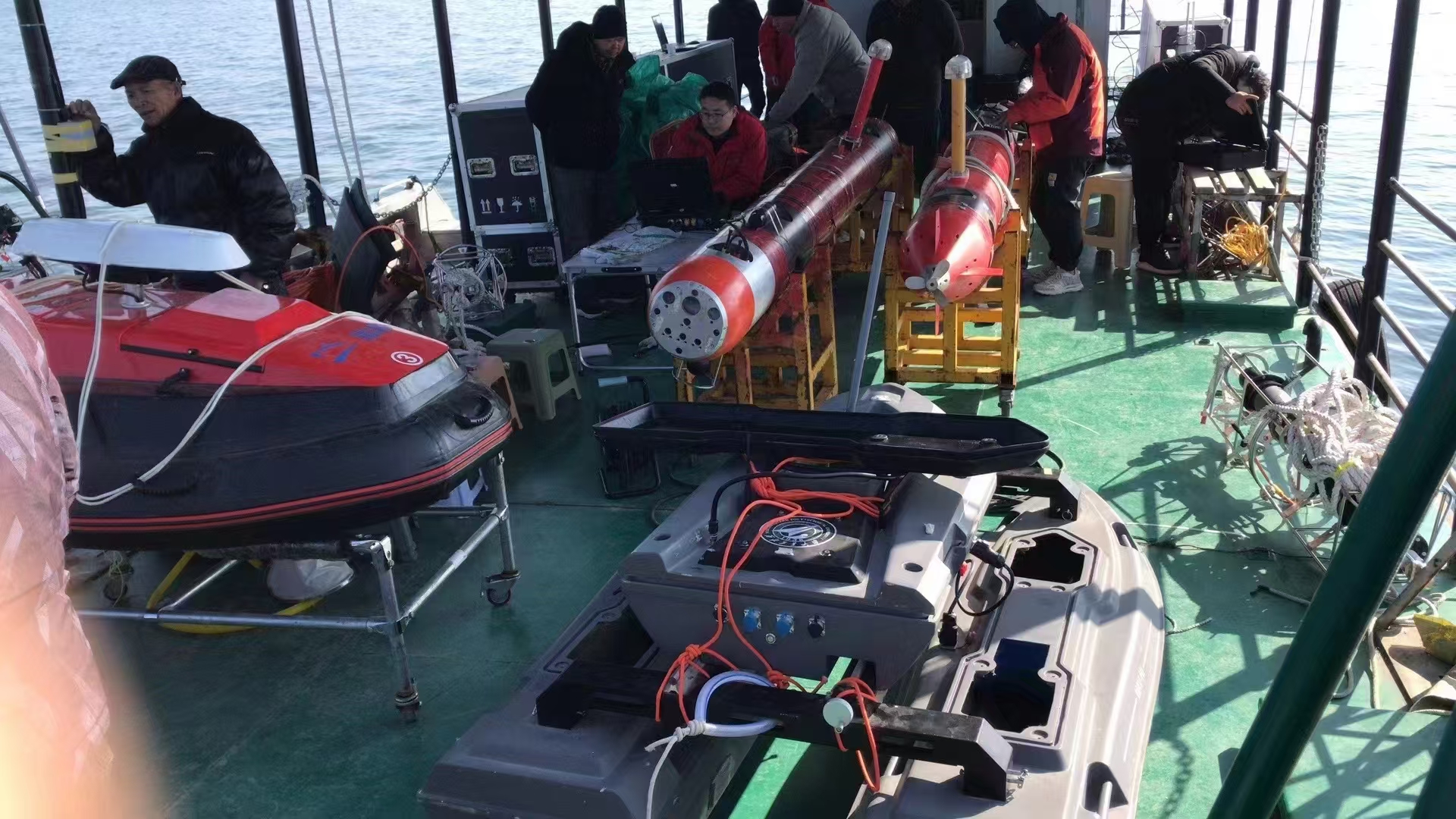}}	

	\subfloat[Development of the experiment \label{p5.6b}]{\includegraphics[width=8cm]{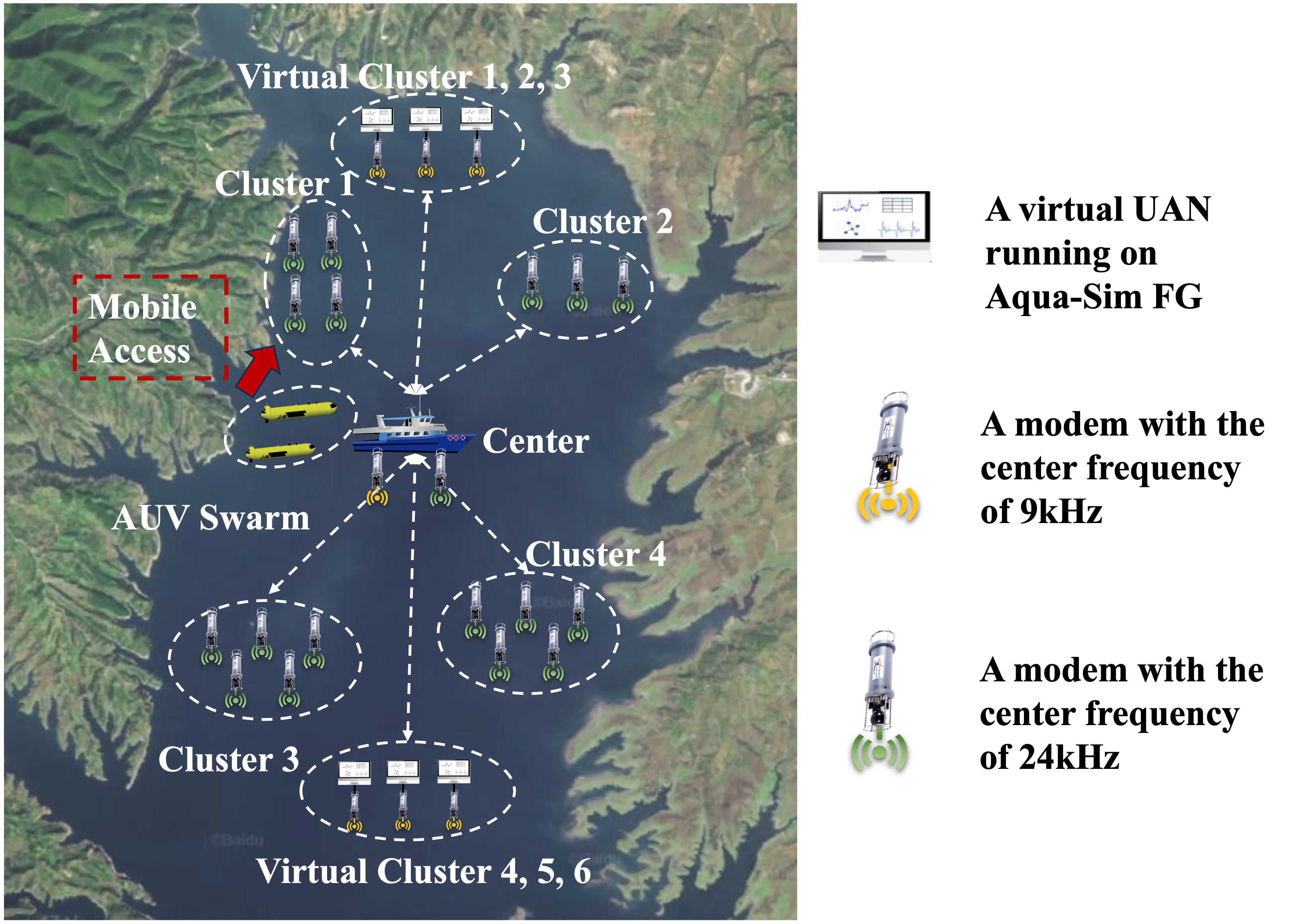}}	

   	\subfloat[Communication between virtual cluster and real modems \label{p5.6c}]{\includegraphics[width=8cm]{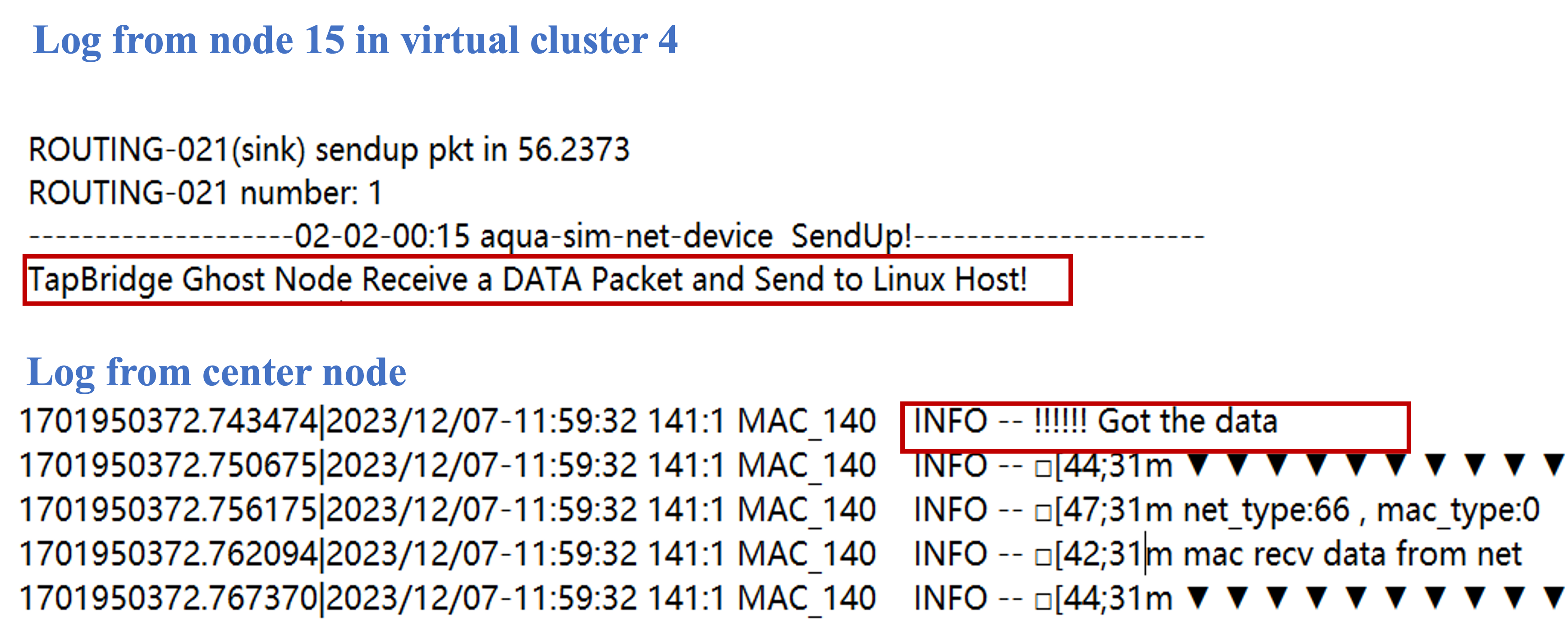}}	
  
	\caption{A large-scale semi-physical network with 140 nodes deployed in Danjiangkou Reservoir, Guangdong, China, 2023. Each virtual cluster is a 20-node virtual UAN running in Aqua-Sim~FG and connects to the real network via a 9~kHz modem.}
	\label{p5.6}	
\end{figure}

\subsubsection{Effects of Header Structure on Transmission Delay} In this set of experiments, we select various protocols (two MAC protocols and one routing protocol) to analyze header structures of Aqua-Sim~FG and previous simulators. As represented in Table.~\ref{t1}, Geo-rOuting Aware MAC protocoL (GOAL) \cite{r25}, SFAMA protocol, and VBF routing protocol are three different protocols, which all have various types of packets. These packets have different functions and include different information. Take GOAL protocols' packets as an example, Aqua-Sim~FG design adaptive headers for request (REQ), reply (REP), data, and acknowledgment (ACK) packets. In this way, Aqua-Sim~FG simulates different transmission delay of these packets’ headers. In previous generations of simulators (take Aqua-Sim~TG as an example), all types of packets adopt a fixed header structure. No matter what type of packets the GOAL protocol transmits, the header size is always 92 B. In such a condition, packets with less header information, like ACK, still spend 1.47 s (the transmission rate is 1500~bps) to transmit header information. Compared with previous simulators, Aqua-Sim~FG provides a more flexible header structure, reducing storage resources and transmission delay. 

\subsubsection{Multiple Mobility Models} In this set of experiments, we simulate an AUV, a UG, and a WG in Aqua-Sim~FG and display their movement trajectories in Fig.~\ref{p5.5}. We design a default path for an AUV at a depth of 40~m, and Aqua-Sim~FG can simulate the AUV's location at different times based on the path. We set that a UG can move in the depth range of 10~m and 30~m, and Aqua-Sim~FG can simulate its zigzag gliding trajectory. For a WG, we adopt a wave model in \cite{r35}, and Aqua-Sim~FG can simulate the trajectory for a WG based on the wave model as represented in Fig.~\ref{p5.5}. By employing these mobility models, users can simulate typical underwater vehicles' movement via Aqua-Sim~FG.

\subsection{A Practical Application Case}

The above experimental results verify that Aqua-Sim~FG can simulate UANs' performance realistically and has high flexibility and usability. We can utilize Aqua-Sim~FG to construct large UANs in a low-cost way. As represented in Fig.~\ref{p5.6}, we deploy a large-scale UAN in Danjiangkou Reservoir, Guangdong, China, 2023. The UAN includes 20 real nodes (AUVs and submersible buoys) and 120 virtual nodes (simulated by Aqua-Sim~FG). Aqua-Sim~FG retains the ability of Aqua-Sim~TG to communicate with real nodes as shown in Fig.~\ref{p5.6c}. In this way, Aqua-Sim~FG overhears the communication rules of real UAN and obtains marine environment information from real nodes. Based on this information, Aqua-Sim~FG adjusts related parameters to simulate virtual clusters more authentically. Under such a condition, virtual clusters and real nodes run on the same network, assisting users in exploring large-scale UANs' performance with low costs.                

\section{Conclusion}

In this paper, we propose the fourth-generation ns-3-based simulator Aqua-Sim~FG for UANs, improving underwater simulation ability to keep abreast of advances in UANs. We refer to the functions of previous generations and design a new general architecture for Aqua-Sim~FG to be compatible with various programming languages, including MATLAB, C++, and Python. In this way, users can employ Aqua-Sim~FG to simulate communication technology, network protocols, and AI models simultaneously in a unified environment. Moreover, we expand new node and communication features, involving various mobility models, adaptive header structure, simple cross-layer interaction, subcarrier-level frequency spectrum configuration method, multiple modulation models, and different accuracy propagation models, to enhance the UAN simulations’ flexibility and fineness. By comparing results of Aqua-Sim~FG and real field experiments, we verify that Aqua-Sim~FG can simulate UANs performance realistically. Extensive simulation results also demonstrate that Aqua-Sim~FG can reflect intelligent methods' problems in real-ocean scenarios, and provide more effective troubleshooting and optimization for actual UANs.

\section*{Acknowledgments}
This work was supported in part by the National Key Research and Development Program of China under Grant 2021YFC2803000; in part by the National Natural Science Foundation of China under Grant 62101211, and Grant 62471201.

\bibliographystyle{IEEEtran}
\bibliography{ref}

\begin{IEEEbiography}[{\includegraphics[width=1in,height=1.25in,clip,keepaspectratio]{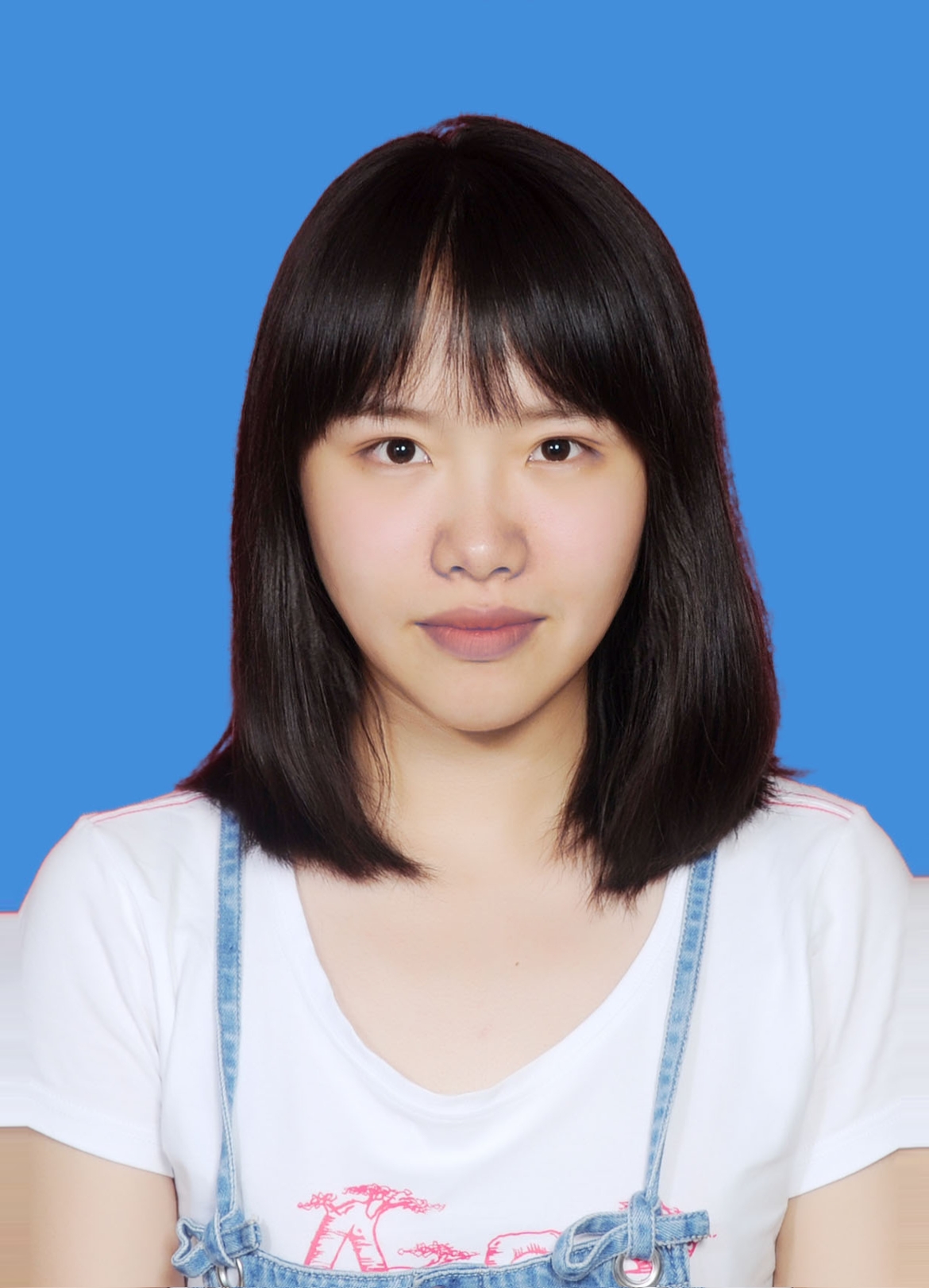}}]{Jiani Guo}
	received the BS degree (2016) in computer science and technology from Beijing Jiaotong University, Beijing, China, received PhD degree (2024) in computer science and technology from Jilin University, Changchun, China. She is currently a Post-Doctoral Researcher with the Department of Computer Science and Technology, Jilin University, Changchun, China. Her current research interests include protocol design, performance analysis, data collection, and simulation design for underwater acoustic networks.
\end{IEEEbiography}
\begin{IEEEbiography}[{\includegraphics[width=1in,height=1.25in,clip,keepaspectratio]{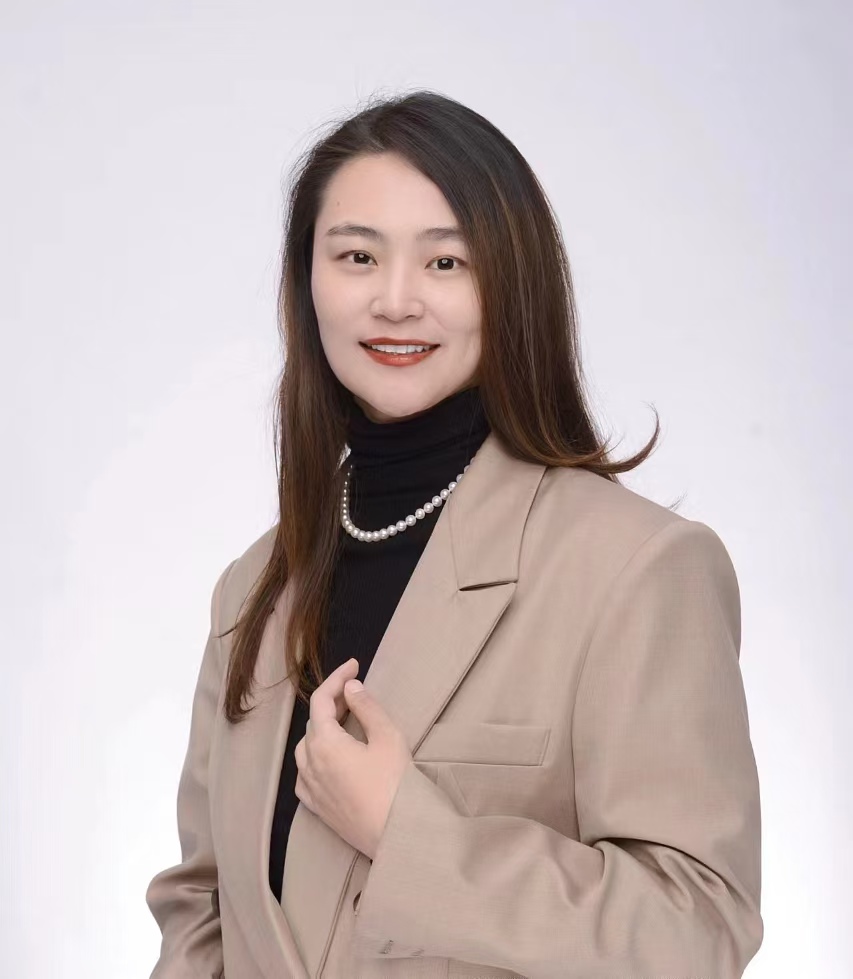}}]{Shanshan Song}
	received the BS degree (2011) and MS degree (2014) in computer science and technology from Jilin University, China, received PhD degree (2018) in Management science and engineering from Jilin University, China. She was a Post-Doctoral Researcher with the Department of Computer Science and Technology, Jilin University, Changchun, China. She is currently an associate professor with the Department of Computer Science and Technology, Jilin University. Her major research focuses on underwater data collection, localization and navigation, and machine learning.
\end{IEEEbiography}
\begin{IEEEbiography}[{\includegraphics[width=1in,height=1.25in,clip,keepaspectratio]{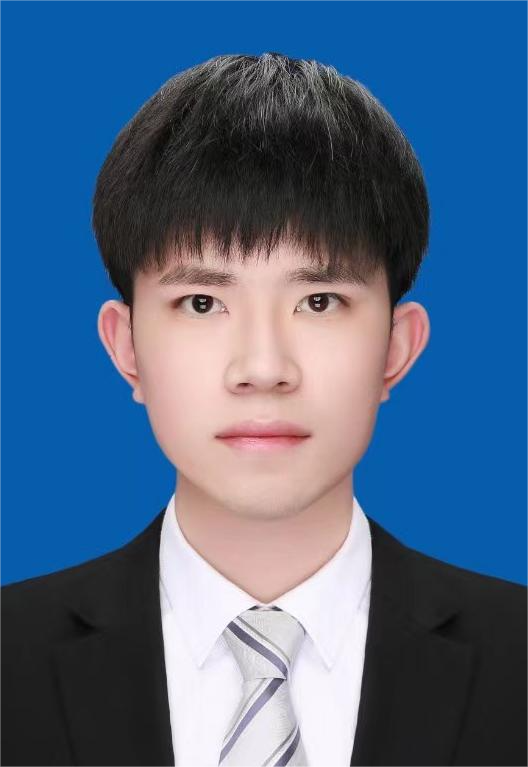}}]{Hao Chen}
	received the BS degree (2022) in computer science and technology from Jilin University, Changchun, China. He is currently working toward the PhD. degree at the College of Computer Science and Technology at Jilin University, Changchun, China. His major research focuses on performance analysis and data collection for underwater acoustic networks.
\end{IEEEbiography}
\begin{IEEEbiography}[{\includegraphics[width=1in,height=1.26
in,clip,keepaspectratio]{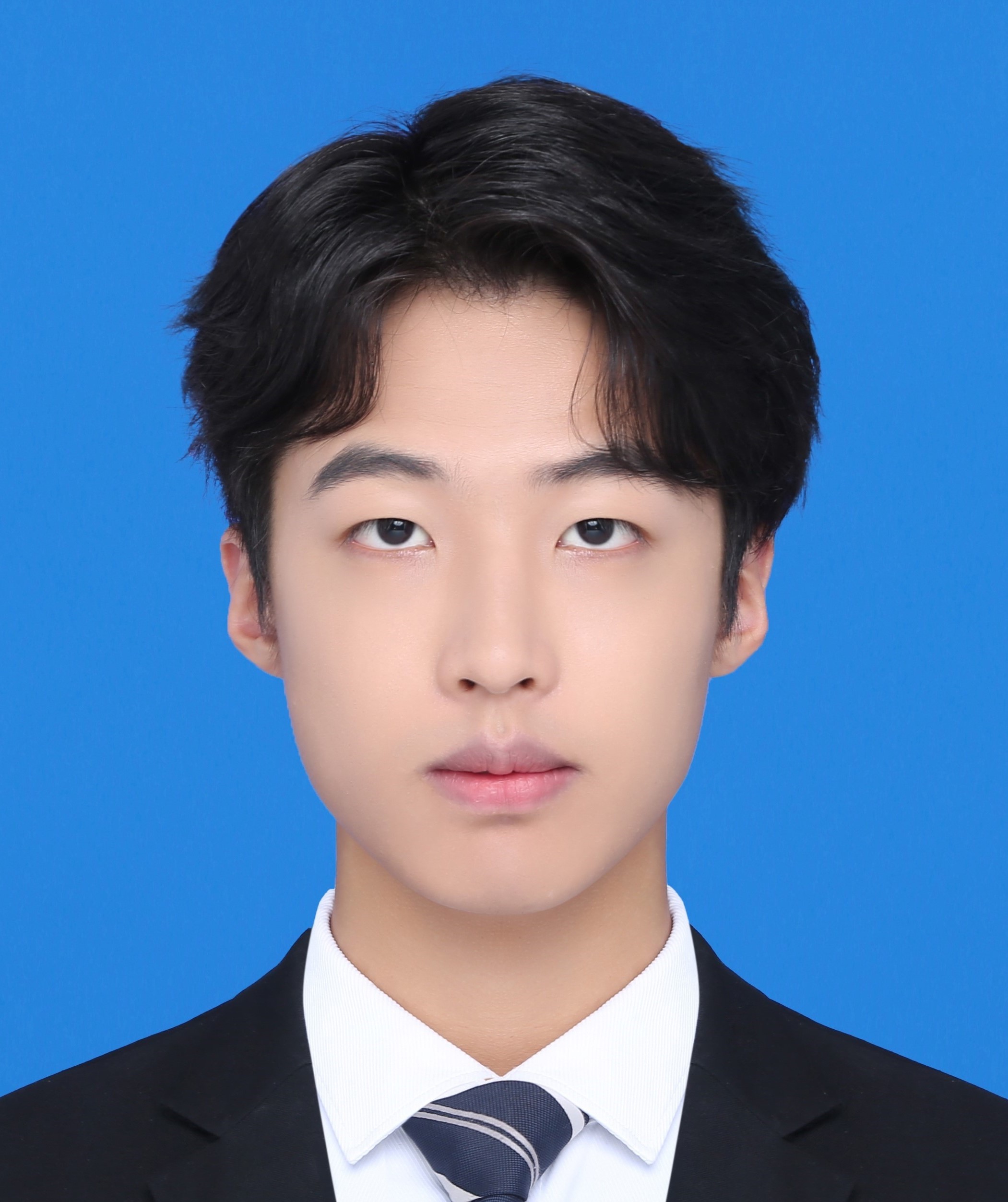}}]{Bingwen Huangfu}
	received the BS degree (2022) in computer science and technology from Jilin University, Changchun, China. He is currently working toward the PhD. degree at the College of Computer Science and Technology at Jilin University, Changchun, China. His major research focuses on digital twin technology for underwater acoustic networks.
\end{IEEEbiography}
\begin{IEEEbiography}[{\includegraphics[width=1in,height=1.25in,clip,keepaspectratio]{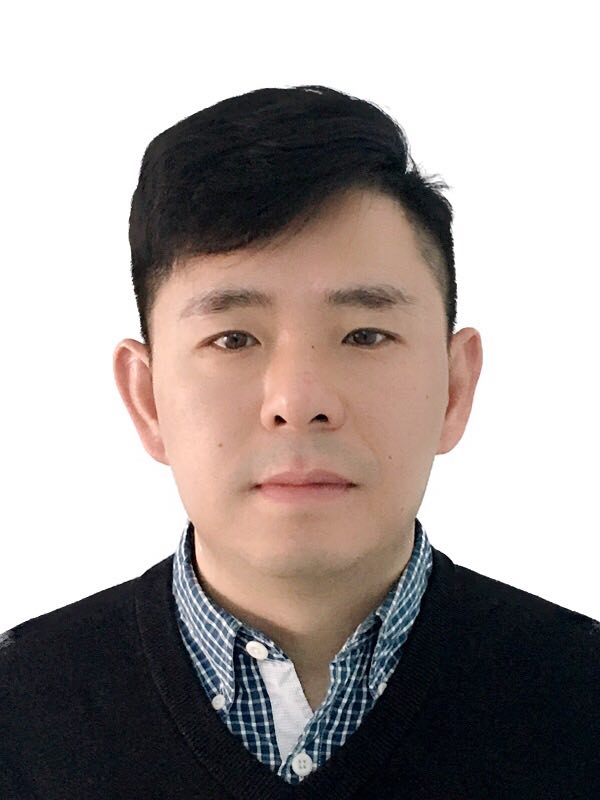}}]{Jun Liu}
	received the BS degree (2002) in computer science from Wuhan University, China, the PhD degree (2013) in Computer Science and Engineering from University of Connecticut, USA. Currently, he is a professor of the School of Electronic and Information Engineering at Beihang University, Beijing, China. His major research focuses on underwater acoustic networking, time synchronization, localization, network deployment, and also interested in operating systems, cross-layer design.
\end{IEEEbiography}

\begin{IEEEbiography}[{\includegraphics[width=1in,height=1.25in,clip,keepaspectratio]{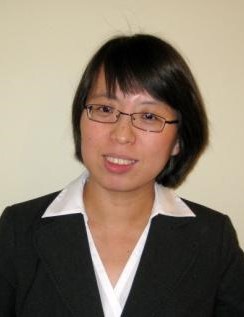}}]{Junhong Cui}
	received the BS degree (1995) in computer science from Jilin University, China, the MS degree (1998) in computer engineering from the Chinese Academy of Sciences, China, and the PhD degree (2003) in computer science from the University of California, Los Angeles. She was on the faculty of the Computer Science and Engineering Department at the University of Connecticut, Storrs. Currently, she is the professor of the College of Computer Science and Technology at Jilin University, Changchun, China. Her research interests include the design, modeling, and performance evaluation of networks and distributed systems. Recently, her research mainly focuses on exploiting the spatial properties in the modeling of network topology, network mobility, and group membership, scalable and efficient communication support in overlay and peer-to-peer networks, and algorithm and protocol design in underwater sensor networks. She is actively involved in the community as an organizer, a TPC member, and a reviewer for many conferences and journals. She is a guest editor for ACM Mobile Computing and Communications Review and Elsevier Ad Hoc Networks. She cofounded the first ACM International Workshop on UnderWater Networks (WUWNet 2006) and now serves as the WUWNet steering committee chair. She is a member of the IEEE, ACM, ACM SIGCOMM, ACM SIGMOBILE, IEEE Computer Society, and IEEE Communications Society.
\end{IEEEbiography}

\vfill

\end{document}